\documentclass[12pt]{article}

\usepackage{amssymb,amsmath,amsfonts,eurosym,geometry,ulem,graphicx,caption,color,setspace,comment,footmisc,caption,natbib,pdflscape,array,hyperref,enumitem,threeparttable,booktabs,tabularx,bbm,pifont,float,graphicx,subcaption,rotating,makecell,multirow,titletoc}

\usepackage{newpxtext,newpxmath}

\usepackage{tikz}
\usetikzlibrary{positioning,matrix,calc,tikzmark}

\newcolumntype{L}[1]{>{\raggedright\let\newline\\arraybackslash\hspace{0pt}}m{#1}}
\newcolumntype{C}[1]{>{\centering\let\newline\\arraybackslash\hspace{0pt}}m{#1}}
\newcolumntype{R}[1]{>{\raggedleft\let\newline\\arraybackslash\hspace{0pt}}m{#1}}

\newtheorem{innercustomhyp}{Hypothesis}
\newenvironment{customhyp}[1]
{\renewcommand\theinnercustomhyp{#1}\innercustomhyp}
{\endinnercustomhyp}

\newtheorem{innercustomres}{Result}
\newenvironment{customres}[1]
{\renewcommand\theinnercustomres{#1}\innercustomres}
{\endinnercustomres}

\begin{document}

\begin{titlepage}
\title{
Testing for Spillovers in Resource Conservation: \\ Evidence from a Natural Field Experiment\thanks{This study was approved by the Departmental Ethics Review Committee of NUS Economics (ECSDERC-2021-10) and pre-registered on AsPredicted (\#76362). Lim gratefully acknowledges support from the Social Science Research Council Graduate Research Fellowship (SSRC-2023-002), administered by the Ministry of Education, Singapore. We thank Rafi Kamsani, Edwin Tan Meng Hong, and Yiwei Fan for excellent research assistance. We are grateful to the NUS Office of Housing Services for providing financial and logistical support in the implementation of the experiment. The views expressed herein are those of the authors and do not reflect the views of the Social Science Research Council (Singapore) and the Ministry of Education, Singapore. All errors that remain are ours.}
}
\author{
	Lorenz Goette\thanks{Department of Economics, National University of Singapore. Email: ecslfg@nus.edu.sg}
    \and 
    Zhi Hao Lim\thanks{Department of Economics, Columbia University. Email: zl2969@columbia.edu.}
    \\[6mm]
	}
\date{\today}
\maketitle

\begin{abstract}
% \noindent
\singlespacing
This paper studies whether behavioral interventions designed to promote resource conservation in one domain generate spillovers in another. Using a natural field experiment involving over 2,000 residents, we identify the direct and spillover effects of real-time feedback and social comparisons on water and energy consumption. We implement three interventions: two targeting shower use and one targeting air-conditioning use. We find significant reductions in shower use from both water-saving interventions, but no direct effect of the energy-saving intervention on air-conditioning use. For spillovers, we estimate precise null effects of water-saving interventions on air-conditioning use, and of the energy-saving intervention on shower use.
\bigskip \\[4mm]
\end{abstract}

\setcounter{page}{0}
\thispagestyle{empty}
\end{titlepage}

% \doublespacing
\onehalfspacing

\section{Introduction}

A wide range of behavioral interventions, such as feedback provision, social comparisons, and moral suasion, have been deployed to influence consumer behavior and promote resource conservation \citep{allcott2011social,allcott2014short,ayres2013evidence,ferraro2011persistence,ferraro2013using,di2017nudging, ito2018moral}. Considerable effort has focused on identifying the direct conservation effects of these interventions within their targeted domains, such as evaluating the impact of home energy reports on energy use or real-time feedback on shower water use. Yet an important question remains largely underexplored: can interventions targeting one resource domain generate spillover effects on behaviors in other, non-targeted domains? Understanding such spillovers is crucial not only for designing comprehensive conservation strategies but also for assessing the welfare implications and cost-effectiveness of behavioral interventions.

Various mechanisms have been proposed to explain behavioral spillovers, often yielding competing predictions (see e.g., \citealp{dolan2015like}). On one hand, positive spillovers could emerge from priming effects, self-identity considerations or preferences for consistency, whereby individuals who engage in pro-environmental behavior in one domain become more likely to do so in another \citep{whitmarsh2010green, van2014looking}. On the other hand, negative spillovers may emerge from moral licensing, whereby individuals feel a sense of moral entitlement after performing well in the targeted domain and subsequently behave less responsibly elsewhere \citep{merritt2010moral, miller2010psychological}. Negative spillovers may also arise from limited attention if interventions focusing on one behavior divert attention away from other domains \citep{trachtmanndoes, altmann2022interventions, koch2024spillover}.

The recent economics literature has begun to investigate the direction and magnitude of behavioral spillovers in resource conservation, though the findings remain mixed. \cite{tiefenbeck2013better} document an increase in energy consumption in response to weekly feedback on water consumption, suggestive of moral licensing. In contrast, \cite{jessoe2021spillovers} report positive spillovers in energy use during summertime, while \cite{carlsson2021behavioral} find reductions in energy consumption only among water-efficient households from their respective water-saving interventions. Additionally, \cite{goetz2024one} observe no impact on electricity consumption from their hot water-saving intervention but document large positive spillovers to room heating energy use. These studies use periodic feedback or social comparisons targeting the focal behavior. Because such interventions often generate relatively modest direct conservation effects, the scope for spillovers may likewise be limited.

This paper investigates behavioral spillovers in resource conservation by examining two key domains: shower use and air-conditioning use, which account for a significant share of water and energy consumption, respectively. Central to our investigation is the deployment of a new class of digital technologies for sustainability---namely, real-time feedback enabled by the Internet of Things. In our setting, smart shower heads provide users with instantaneous feedback on their water usage. This approach has been shown to be more effective in promoting conservation than traditional interventions based on social comparisons and moral suasion \citep{tiefenbeck2018overcoming, fang2023complementarities, goette2021dynamics}, thereby increasing the potential for larger spillover effects. Moreover, these real-time interventions may operate through an additional behavioral channel---limited attention---creating the possibility of attentional spillovers alongside mechanisms such as moral licensing or norm adherence \citep{medina2021side, altmann2022interventions, trachtmanndoes, koch2024spillover}. 

We conduct a randomized field experiment with approximately 2,200 residents across four university residences, implementing three interventions: two targeting shower use and one targeting air-conditioning use. Specifically, we target shower use through real-time feedback and social comparison interventions, and examine spillover effects on air-conditioning use. In parallel, we target air-conditioning use through social comparisons and examine spillovers to shower use. This design allows us to study cross-domain spillover effects from water use to energy use and vice versa, extending the existing literature, which typically considers spillovers in only one direction. Our large sample size enables us to precisely estimate spillover effects and examine patterns that shed light on potential mechanisms, including attentional spillovers and moral licensing. In addition, this study qualifies as a natural field experiment, as residents were unaware of their participation and the interventions were embedded within a broader university-led resource conservation initiative \citep{harrison2004field}. Together, our findings provide new evidence on the extent of behavioral spillovers and highlight both the potential and limits of digital technologies for promoting resource conservation. In particular, we present three main sets of results.

First, we find significant reductions in shower water use from both water-saving interventions: real-time feedback reduces usage by 15.9\%, while social comparisons generate more modest reductions of 3.5\%. In contrast, the energy-saving intervention has no direct effect on air-conditioning use. Second, we find no evidence of cross-domain spillovers, with precisely estimated null effects in both directions: the water-saving interventions do not affect air-conditioning use, and the energy-saving intervention does not affect shower water use. Third, we find evidence of within-domain heterogeneity in the showering domain, where treatment effects of social comparisons decrease with relative performance ranking, but not in the air-conditioning domain or across domains.

The rest of the paper proceeds as follows. Section 2 describes the experimental design and hypotheses. Section 3 presents the results. Section 4 discusses implications of our findings and concludes.
\section{Design and Hypotheses} 

The field experiment involved approximately 2,200 residents across four residential colleges at NUS University Town in Singapore and was conducted from August 9, 2021 to December 4, 2021. Two classes of behavioral interventions, real-time feedback and social comparisons, were employed to assess the direct and spillover effects on residents' resource use behaviors. We pre-registered our design and analysis plan on AsPredicted (\#76362).

\subsection{Experimental Design} 

\begin{figure}[ht]
    \vspace{0.4cm}
    \centering
    \caption{Overview of the Experimental Design}
    \label{fig:diagram-poster}

\begin{tikzpicture}[font=\small]
\def\TextBoxWidth{2.2cm}
\def\HorGapA{2mm}
\def\HorGapB{2mm}
\def\HorGapC{2mm}
\def\VerGapA{5mm}
\def\VerGapB{3mm}
\def\VerGapC{14mm}
\def\VerGapD{2mm}
\def\MainTextBoxHeight{4.5cm}
\pgfmathparse{0.5*(\MainTextBoxHeight-\VerGapA)}
\edef\SmallTextBoxHeight{\pgfmathresult pt}

\begin{scope}[line width=0.25mm, every node/.style={draw,minimum width=\TextBoxWidth,align=center,}]

\node (Main Text Box)
  [minimum height=\MainTextBoxHeight]
  {Control};

\node (Small Text Box 1)
  [minimum height=\SmallTextBoxHeight,below right=0pt and \HorGapA]
  at(Main Text Box.north east)
  {Real-time\\ feedback \\(RTF)};

\node (Small Text Box 2)
  [minimum height=\SmallTextBoxHeight,above right=0pt and \HorGapA]
  at(Main Text Box.south east)
  {Control};
\end{scope}

\begin{scope}[line width=0.25mm, every node/.style={
  draw,
  minimum width=\TextBoxWidth,
  minimum height=0.3333*\SmallTextBoxHeight,
  inner sep=0pt,outer sep=0pt,
  align=center,}]

% Control
\node (Small Text Box 3)
  [below right=0pt and \HorGapB]
  at(Small Text Box 1.north east) {RTF};

% SC-S
\node (Small Text Box 4)
  [below right]
  at(Small Text Box 3.south west) {RTF+SC-S};

% SC-A
\node (Small Text Box 5)
  [below right]
  at(Small Text Box 4.south west) {RTF+SC-A};

\draw (Small Text Box 3.north west) rectangle (Small Text Box 5.south east);

% RTF
\node (Small Text Box 6)
  [below right=0pt and \HorGapB]
  at(Small Text Box 2.north east) {Control};

% RTF+SC-S
\node (Small Text Box 7)
  [below right]
  at(Small Text Box 6.south west) {SC-S};

% RTF+SC-A
\node (Small Text Box 8)
  [below right]
  at(Small Text Box 7.south west)  {SC-A};

\draw (Small Text Box 6.north west) rectangle (Small Text Box 8.south east);
\end{scope}

\path (Main Text Box.north west);
\pgfgetlastxy{\MainTextBoxNWx}{\MainTextBoxNWy}
\path (Small Text Box 1.north east);
\pgfgetlastxy{\SmallTextBoxOneNEx}{\SmallTextBoxOneNEy}
\pgfmathparse{\SmallTextBoxOneNEx-\MainTextBoxNWx}
\edef\GraphOneWidth{\pgfmathresult pt}

\node (Graph 1) [
  minimum width=\GraphOneWidth,
  inner sep=0pt,
  above right=\VerGapB and 0pt]
  at(Main Text Box.north west)
  {\includegraphics[height = 4.7cm]{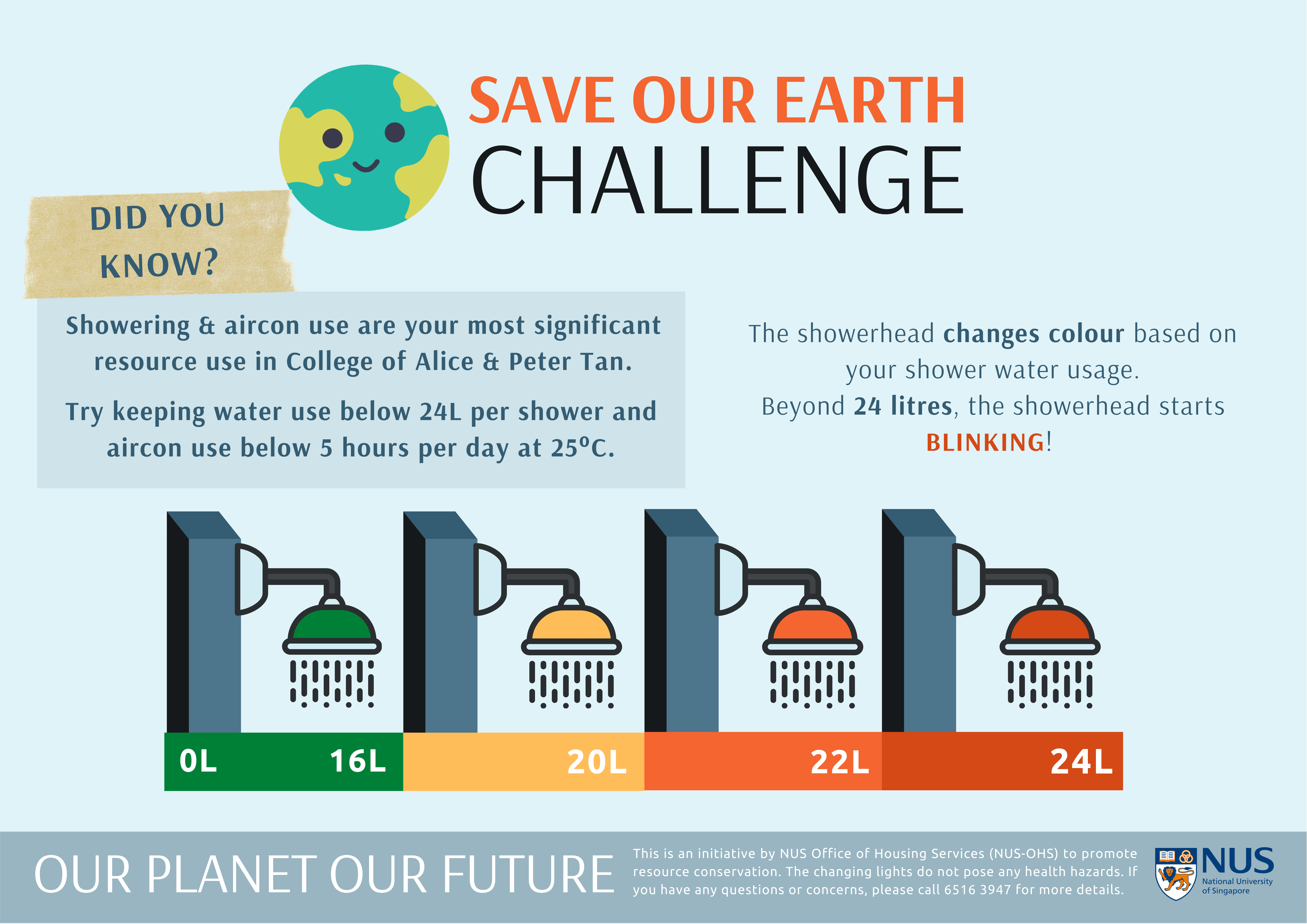}};

\node (Graph 2) [
  inner sep=0pt,
  above right=\VerGapB and 100pt]
  at(Small Text Box 3.north west)
  {\includegraphics[height =4.7cm]{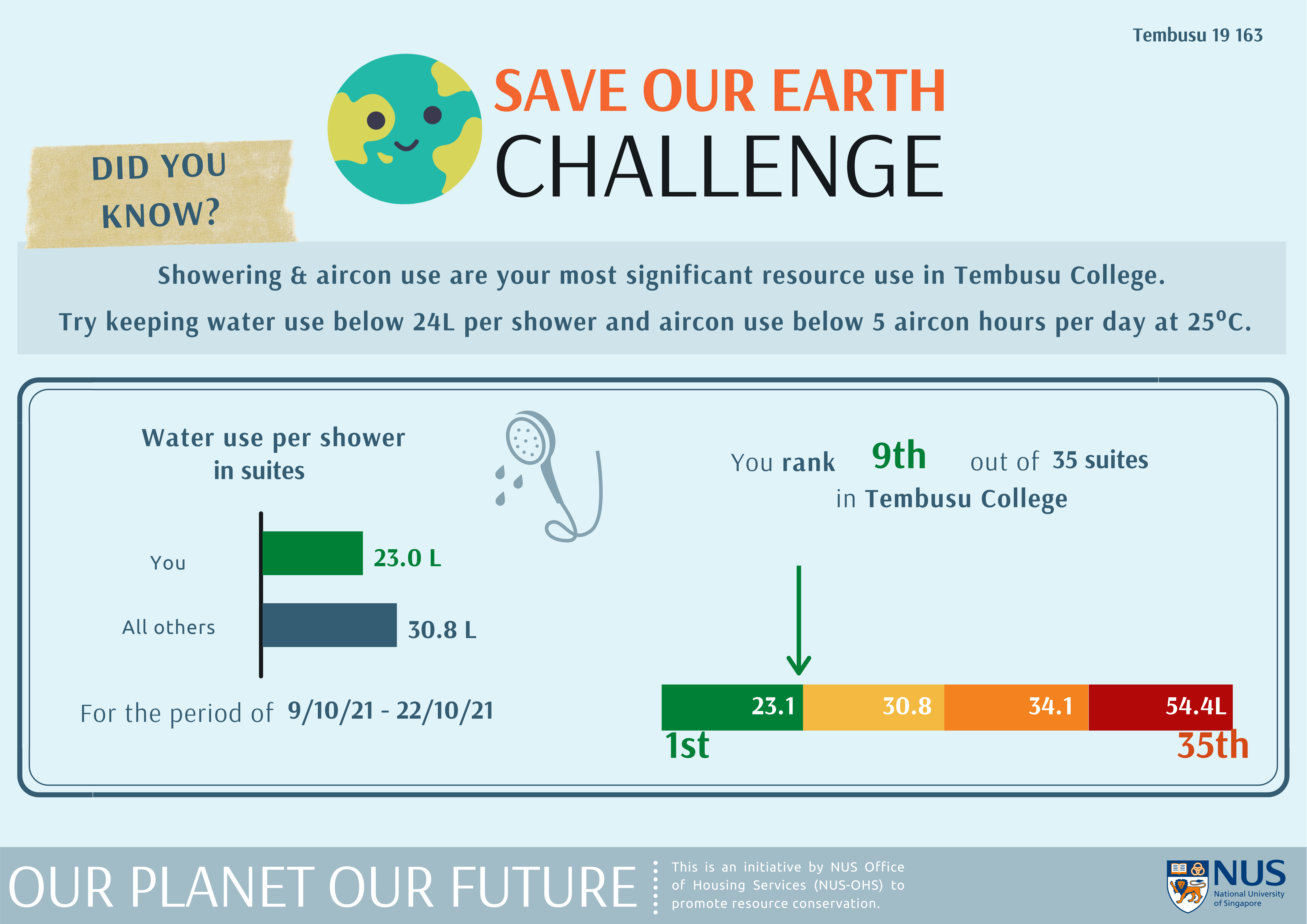}};

\node (Graph 5) [inner sep=0pt,below=\VerGapB]
  at(Graph 2.south)
  {\includegraphics[height =4.7cm]{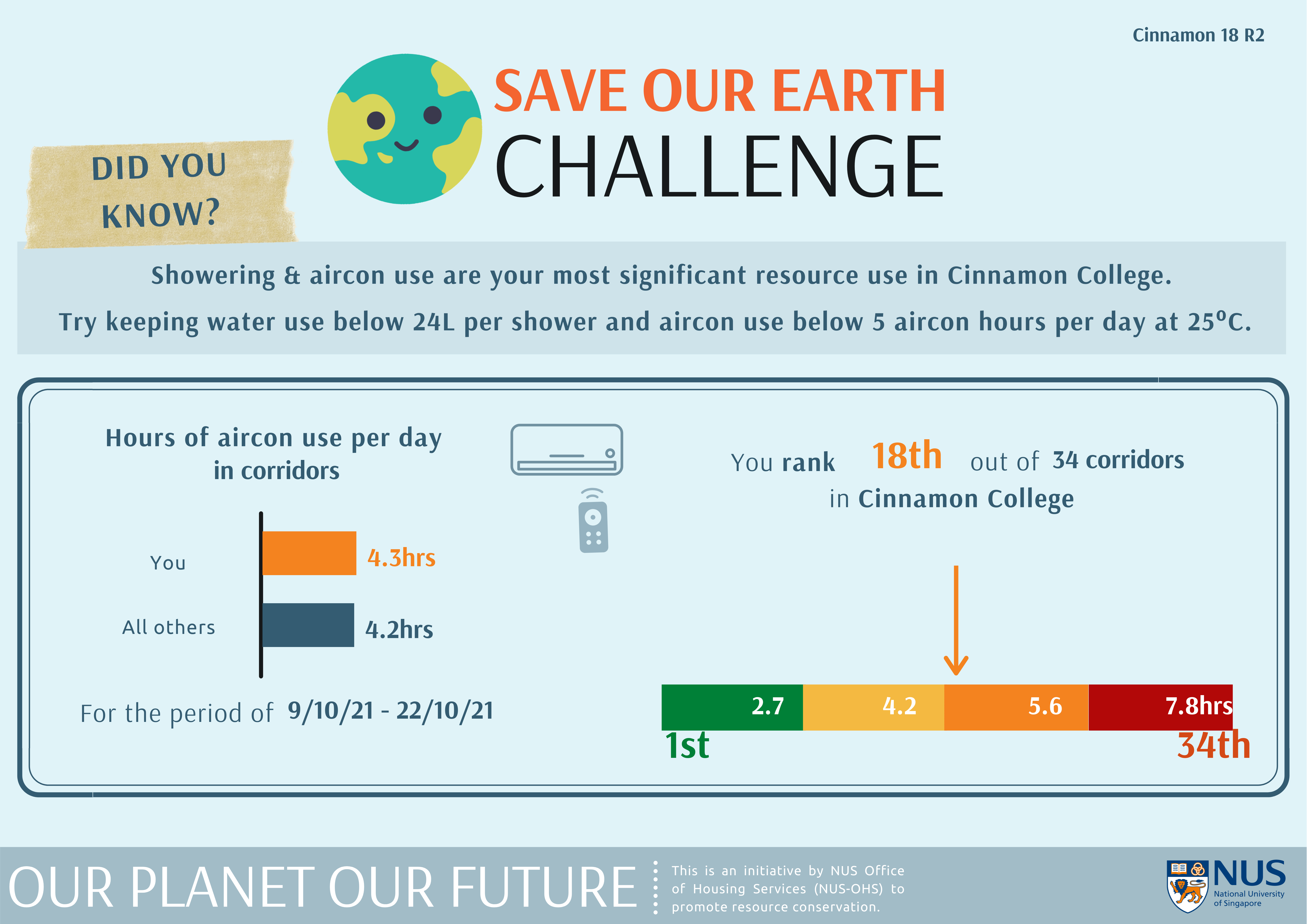}};

 \node (Axis label Time) [below left=\VerGapC and 10pt] at(Small Text Box 6.south east -| Graph 5.west) {};

\coordinate (Ref Point 0) at(Main Text Box.south west);

\coordinate (Ref Point 1) at($(Main Text Box.south east)!0.5!(Main Text Box.south east -| Small Text Box 2.south west)$);

\coordinate (Ref Point 2) at($(Small Text Box 2.south east)!0.5!(Small Text Box 2.south east -| Small Text Box 6.south west)$);

\coordinate (Ref Point 3) at(Small Text Box 6.south east);

\coordinate (Ref Point 4) at($(Small Text Box 6.south east)!0.5!(Small Text Box 6.south east -| Graph 5.south west)$);

\coordinate (Ref Point 5) at($(Small Text Box 6.north east)!0.5!(Small Text Box 6.north east |- Small Text Box 5.south east)$);

\coordinate (Ref Point 6) at($(Small Text Box 3.north east)!0.5!(Small Text Box 6.north east |- Graph 1.south east)$);

\begin{scope}[line width=0.25mm,]

\foreach \i in {0,1,2,3}
{
  \draw (Ref Point \i |- Axis label Time) coordinate(Axis Point \i) +(0,-1mm)--+(0,1mm);
}

\foreach \i/\l [remember=\i as \j (initially 0)] in {1/Baseline,2/Phase 1,3/Phase 2}
{
  \draw (Axis Point \j)--node[below]{\l}(Axis Point \i);
}
\draw [->](Axis Point 3)--(Axis label Time);
\end{scope}

\begin{scope}[line width=0.3mm, draw=black,-,shorten >=1pt]
\draw (Small Text Box 3.north)-- (Graph 1.south -|Small Text Box 3.north);
 \draw (Small Text Box 7.east) -- (Graph 2.west);
 \draw (Small Text Box 8.east)--(Small Text Box 8.east -| Graph 5.west);
\end{scope}
\end{tikzpicture}

\vspace{0.8cm}

\footnotesize \parbox{\linewidth}{\textit{Notes:} The figure outlines our experimental design, where `SC-S' denotes social comparison for shower usage and `SC-A' denotes social comparison for air-conditioning usage. The RTF intervention features a poster (top left) explaining the color changes of the shower head as water usage increases. The SC-S intervention features a poster (top right) comparing average shower water usage of the assigned bathroom with that of other bathrooms in the same residential college and room type. Similarly, the SC-A intervention features a poster (bottom right) comparing the average air-conditioning usage of rooms linked to the assigned bathroom with that of rooms linked to other bathrooms in the same residential college and room type. For examples of each poster type across all six experimental conditions, refer to Figure \ref{fig:sample_posters}.}
\end{figure}

Our experimental design, presented in Figure \ref{fig:diagram-poster}, consists of six experimental conditions built around three interventions: real-time feedback for shower usage (RTF), social comparison for shower usage (SC-S), and social comparison for air-conditioning usage (SC-A). Each intervention is described as follows:

\vspace{3mm}
\noindent
\textbf{RTF intervention:} Residents receive individualized real-time feedback on their own water usage during showers. Within each shower facility, a smart shower head equipped with built-in LEDs is programmed to provide instantaneous feedback on water consumption through color changes. At the start of a shower, the smart shower head displays a green light, which changes sequentially to yellow, orange, and then red as water usage increases (Figure \ref{fig: rtf_treatment}). If water consumption exceeds 24 litres, the shower head signals this with a blinking red light. The color change thresholds are clearly explained via a detailed poster displayed prominently in the assigned shower facility (Panel D of Figure \ref{fig:sample_posters}).

\vspace{3mm}
\noindent
\textbf{SC-S intervention:} Residents receive comparative feedback on shower water usage for the bathroom they use in their corridor or apartment suite. Specifically, the poster reports the average shower water usage of residents sharing that bathroom over the preceding two weeks, and compares it with the average usage of other bathrooms in the same residential college and room type. Bathrooms are then ranked according to these average usage levels, from 1 for the lowest-usage bathroom to a maximum of 37 for the highest-usage bathroom. These rankings are grouped into four quartiles and displayed using a color scale, from green (lowest quartile) to red (highest quartile). The poster is displayed prominently in the assigned shower facility (Panel B of Figure \ref{fig:sample_posters}).

\vspace{3mm}
\noindent
\textbf{SC-A intervention:} Mirroring the SC-S intervention, residents receive comparative feedback on the average daily air-conditioning usage of the rooms linked to the bathroom they use, relative to other bathrooms in the same residential college and room type. Bathrooms are similarly ranked according to these average usage levels, from 1 for the lowest-usage bathroom to 41 for the highest-usage bathroom, with rankings displayed using the same color-coded quartile scheme (Panel C of Figure \ref{fig:sample_posters}).

% \documentclass[tikz]{standalone}
% %\documentclass{ctexart}
% %\usepackage{tikz}
% \usetikzlibrary{positioning,matrix,calc,shapes.multipart}

Both types of social comparison posters are framed around the bathroom residents use relative to other bathrooms in the same residential college and room type (e.g., ``You rank 9th out of 35 suites in Tembusu College''). While we cannot directly verify whether residents fully interpreted the displayed ranking as reflecting shared bathroom-level usage, it is likely that they understood the comparison to be at the group level, since suite and corridor classifications are familiar to residents in this setting.

% Address concerns that all interventions are compared to a control group that is nudged on both behaviors via a similar poster
All the interventions are compared to a control group that receives only a generic poster encouraging residents to conserve resources, without any real-time feedback or social comparison. One concern is that such a poster constitutes a light-touch nudge that may influence behavior. If so, our measured effects may understate the full impact of the interventions relative to a ``pure control'' group that receives no messaging at all. However, we note that this concern is likely minimal, as a prior study in a similar residential college setting finds that conservation posters have no significant impact on resource use \citep{goette2021dynamics}.\footnote{In that study, residents in the \textit{Moral Suasion} group received a poster encouraging water conservation in the shower, which parallels our control group. The estimated effect, relative to a pure control group with no poster, was not significant.}

These interventions were rolled out in phases. The baseline phase involved no interventions, allowing us to confirm balance across experimental conditions (see Appendix Table \ref{tab:sum_stats}). 
In phase 1, half of the residents served as the control group, while the other half received the RTF intervention. In phase 2, the social comparison interventions were introduced across both groups. Specifically, within the control (RTF) group from phase 1, one-third were assigned to the SC-S intervention, another third to the SC-A intervention, and the remaining third continued without any intervention (with only RTF intervention).

% Discuss unit of randomization
Each residential college offers two room types: single corridor rooms and single rooms within shared 6-bedroom apartment suites. On each floor, residents in single corridor rooms share a common bathroom, while those in apartment suites share a private bathroom with other suite occupants (see Figure \ref{fig:floor_plan} for the floor plan).\footnote{Rental rates for a given room type are the same across residential colleges and floors. Apartment-suite rooms are more expensive than corridor rooms, with a price difference of about 7\%.} Because shower water use is measured through smart shower heads installed within bathrooms, treatment is randomized at the bathroom level. Each bathroom contains two shower facilities on average, and when a bathroom is assigned to a treatment arm, all shower heads in that bathroom and the rooms whose residents use that bathroom are assigned to the same intervention. Across the four residential colleges, our study covers 257 bathrooms (clusters) in total, with 481 smart shower heads in use. It also encompasses 2,210 occupied rooms, of which 1,276 have an air-conditioning unit used in our air-conditioning analysis. Table \ref{tab:sum_stats} in the Appendix provides a detailed breakdown of bathrooms (clusters) and bedrooms by experimental condition.

% Discuss possibility of sorting across treated & control bathrooms
A potential concern is sorting across bathrooms. In our setting, such sorting is likely limited because residents typically store their toiletries in a specific bathroom and use that same facility for showers. We further assess this possibility by testing whether treatment effects differ between single-gender and mixed-gender floors, where the cost of switching bathrooms differs, and find no statistically significant differences (see Figure \ref{fig:effects_floorType}).

% Discuss attribution problem, free-riding, and split incentives problem
We also note that a distinction of our design is that the RTF intervention provides individual-level feedback on residents' own water usage, whereas the SC-S and SC-A interventions provide group-level information based on bathroom-level averages. With group-level feedback under the social comparison interventions, residents cannot fully attribute observed performance rankings to their own resource usage. This may give rise to free riding (e.g., \citealp{elinder2017consequences}), as each resident's conservation efforts contribute to a shared outcome, potentially weakening incentives to conserve. In addition, split incentive considerations may also be relevant in our setting \citep{gillingham2012split, jessoe2020utilities}. Water usage is covered by the monthly rent, so residents face zero marginal cost of consumption, which may further weaken incentives to conserve. Air-conditioning usage, however, is priced on a pay-per-use basis, so such concerns are less relevant in that domain. Notwithstanding, these considerations would tend to attenuate behavioral responses to our treatments, particularly for the SC-S intervention, implying that our estimated treatment effects may be conservative in magnitude.

% Discuss COVID-19 context and external validity
Finally, we note that the field experiment was conducted during the COVID-19 pandemic. Prior to the pandemic, the four residential colleges could accommodate up to 2,400 students per semester. During our study period, room availability was reduced by approximately 10\% by the NUS Office of Housing Services to comply with safe management measures introduced by the Singapore government. Despite this reduction, on-campus occupancy remained high (above 90\%), with a total of 2,210 occupied rooms. While pandemic-related restrictions may have affected overall resource use or behavioral responsiveness (for example, if residents spent more time indoors), we note that a prior study conducted in two of the same residential colleges before COVID-19 finds that real-time feedback induces effect sizes on shower water usage that are comparable to those observed in our study \citep{goette2021dynamics}. This strengthens the external validity of our findings beyond the COVID-19 context.

\subsection{Outcomes of Interest}

To study resource use behavior, our analysis focuses on two primary outcome variables: shower water usage and air-conditioning energy usage, which together account for the most substantial shares of residents' overall water and energy consumption.\footnote{One potential asymmetry in our design is that showering involves both water consumption and some energy consumption for water heating, whereas air conditioning involves energy consumption only. In Singapore's tropical climate, however, ambient water temperature is typically around 24°C, and desired shower temperatures are on average lower than in colder settings, so the energy required for shower heating is less substantial relative to air conditioning. Therefore, we view our interventions as primarily targeting two distinct resource domains: the RTF and SC-S interventions target water consumption, while the SC-A intervention targets energy consumption.} Data on shower water usage were collected using HYDRAO smart shower heads installed in each shower facility, which automatically record water usage for each shower event and transmit the data via WiFi to a central server.

Data on air-conditioning energy usage were obtained from Power Automation, the company which manages the metering systems in the residential colleges. Air-conditioning usage is metered at the individual bedroom level through the university's pay-as-you-use system. The raw data consist of cumulative electricity consumption from air-conditioning use (in kilowatt-hours, kWh), which we convert into daily usage (in hours) using institutionally provided cost and efficiency rate parameters. All bedrooms are single-occupancy, so the air-conditioning outcome reflects individual residents’ usage patterns.

In rooms equipped with air conditioning, residents must actively choose usage duration on a pay-per-use basis and ensure that they have sufficient account credit.\footnote{Residents can top up their accounts online or at a physical terminal in the residence management office. See \url{https://nus-utown.evs.com.sg/}} Each time they turn on the air-conditioning unit, residents need to choose the number of active minutes of cooling airflow, but cannot adjust the thermostat to a specific temperature setting. Common areas in shared apartment suites are not air-conditioned and are therefore excluded from our measure.

For each resource type, we distinguish between intensive and extensive margins of usage. For showers, the intensive margin is the average volume per shower (in litres), while the extensive margin is the daily number of showers per shower head. For air conditioning, the intensive margin is total daily usage (in hours), and the extensive margin captures whether the air-conditioning unit was used on a given day.

\subsection{Empirical Predictions}

The literature on behavioral spillovers proposes several mechanisms through which interventions targeting one behavior may affect other, non-targeted behaviors. We discuss potential mechanisms that could generate spillover effects in our setting and outline the empirical predictions pertaining to our interventions.

One potential mechanism is attentional spillovers, which relate to the literature on inattention in decision-making \citep{gabaix2014sparsity, gabaix2019behavioral}. Interventions targeting behavior in one domain may shift individuals' attention towards that activity, altering how limited attention is allocated across other domains. \cite{nafziger2020spillover} formalizes a framework in which interventions that increase attention to a particular domain may generate spillovers depending on whether the targeted and non-targeted domains are attentional substitutes, complements, or independent.

In our setting, the RTF intervention likely increases the salience of shower water use. The smart shower head uses a color-changing light to signal water usage, with a blinking red light indicating excess usage. By making water consumption visually salient during the shower, the intervention may draw residents' attention to their showering behavior. If showering and air-conditioning usage compete for the user's limited attention, directing attention towards showering may inadvertently divert attention away from air-conditioning usage, potentially generating negative spillovers. Evidence consistent with such negative cross-domain spillovers has been documented in other contexts. For example, \cite{trachtmanndoes} shows that messages promoting one healthy behavior (meal-logging) reduced completion rates in another non-targeted behavior (meditation), and vice versa. \cite{medina2021side} finds that reminders for upcoming credit card payments increased overdraft fees in checking accounts, while \citet{altmann2022interventions} document negative cognitive spillovers impacting decision quality in domains not targeted by their intervention.

Another possible mechanism is thermal comfort substitution. Showering and air conditioning both provide heat relief and affect thermal comfort levels \citep{salvo2018electrical}. If residents reduce the duration of their showers in response to the RTF intervention, they may compensate by increasing air-conditioning usage to maintain a similar level of thermal comfort. In this case, reductions in shower water use could be accompanied by increases in air-conditioning usage, generating the same empirical prediction as attentional spillovers when the two domains are attentional substitutes. This motivates our first hypothesis:

\begin{customhyp}{1A:}{\label{hypo: 1a}}
\textbf{Negative cross-domain spillovers.} The RTF intervention reduces shower water usage (targeted behavior), but leads to an increase in air-conditioning usage (non-targeted behavior).
\end{customhyp}

Alternatively, if showering and air-conditioning usage are attentional complements, directing attention toward showering may increase awareness of resource use more broadly. In this case, the RTF intervention may also increase attention to air-conditioning usage, potentially leading residents to reduce energy consumption in that domain as well. Evidence consistent with positive attentional spillovers has been documented in other contexts. For example, \cite{simonov2023attention} find that engaging news articles increased attention to adjacent advertisements on the same page in an online setting. 

Beyond attentional factors, positive spillovers may also arise through information effects, whereby individuals process the information they receive and become more aware of their broader resource use patterns. \cite{barlose2024behavioral} provide evidence supporting this channel in the context of promoting healthier food choices. This motivates a competing hypothesis:

\begin{customhyp}{1B:}{\label{hypo: 1b}}
\textbf{Positive cross-domain spillovers.} The RTF intervention reduces shower water usage (targeted behavior), but also leads to a decrease in air-conditioning usage (non-targeted behavior).
\end{customhyp}

Another potential mechanism is moral licensing, whereby engaging in morally commendable actions may confer a sense of ``moral credit,'' leading individuals to feel justified in subsequently engaging in less responsible behavior \citep{merritt2010moral,  miller2010psychological}. This phenomenon has been documented across a wide range of domains, including racial attitudes \citep{effron2009endorsing}, consumer choice \citep{khan2006licensing}, health behaviors \citep{chiou2011randomized}, and charitable giving \citep{meijers2015dark}.\footnote{See \cite{blanken2015meta} for an extensive review.} 

While moral licensing can arise within the same domain, it may also occur across domains. In the context of resource consumption, engaging in pro-environmental actions in one domain may lead individuals to behave less responsibly in another. A study related to ours is \citet{tiefenbeck2013better}, who document suggestive evidence of moral licensing across domains: residents who received weekly feedback on their water consumption reduced water use but concurrently increased energy use.

In our experiment, the SC-S and SC-A treatments provide residents with information about how their bathroom's average shower usage and air-conditioning usage, respectively, are ranked against other comparable bathrooms. Such relative performance feedback may generate heterogeneous responses depending on whether residents learn they are performing well or poorly. A moral licensing interpretation predicts that residents receiving more favorable relative performance feedback may subsequently conserve less, while those receiving less favorable feedback may conserve more. This could occur within the targeted domain or across domains. This leads to the following hypotheses:

\begin{customhyp}{2:}{\label{hypo: 2}}
\textbf{Within-domain heterogeneity by performance ranking.} In the SC-S (SC-A) intervention, subjects who are ranked higher (i.e., lower resource usage) exhibit smaller reductions in subsequent shower (air-conditioning) usage in the targeted domain.    
\end{customhyp}

\begin{customhyp}{3:}{\label{hypo: 3}}
\textbf{Cross-domain heterogeneity by performance ranking.} In the SC-S (SC-A) intervention, subjects who are ranked higher (i.e., lower resource usage) exhibit higher air-conditioning (shower) usage in the non-targeted domain.
\end{customhyp}

We note that within-domain treatment heterogeneity (Hypothesis 2) is not, by itself, uniquely diagnostic of moral licensing. One alternative mechanism is norm adherence or conformity, whereby residents adjust their consumption behavior toward the perceived social norm (e.g., \citealp{allcott2011social}). Another is conditional cooperation, whereby residents condition their own conservation effort on the contribution of others (e.g., \citealp{frey_social_2004}). These mechanisms, however, do not naturally account for cross-domain responses (Hypothesis 3), since residents are not provided information about consumption behavior in the non-targeted domain. We therefore interpret Hypothesis 3 as more closely tied to the possibility of cross-domain moral licensing.

\subsection{Responsiveness of Resource Usage to External Influences} 

Extensive evidence demonstrates that showering behavior is highly responsive to behavioral interventions, particularly real-time feedback \citep{tiefenbeck2018overcoming, agarwal2022goal, goette2021dynamics, fang2023complementarities}. In contrast, evidence on the responsiveness of air-conditioning usage is more mixed. \citet{goette2021effects} find no overall effect of their social comparison interventions on air-conditioning usage, although reductions are observed among residents with the lowest baseline consumption. By comparison, \citet{tiefenbeck2013better} report that a weekly water feedback intervention increased overall electricity consumption, likely driven by higher air-conditioning usage, which constitutes a large share of household energy demand.

Against this backdrop, we first examine whether showering and air-conditioning usage are responsive to external weather conditions. In our setting, both behaviors are sensitive to daily temperature fluctuations, suggesting that residents can adjust their consumption in meaningful ways. Establishing this baseline responsiveness is useful for assessing both the direct and spillover effects of our interventions, as it shows that residents have scope to change behavior.\footnote{A similar logic is used by \citet{myers2020social}, who find that although their main social comparison-based intervention yielded null effects, students did respond to a simple email reminder to lower thermostats before leaving for the winter break. This demonstrates that participants were attentive to the interventions and capable of adjusting their behavior.}

Table \ref{tab:weather_impacts} presents regression estimates of both resource usage outcomes on daily average temperature during the experimental period. For showering, a $1^{\circ}$C increase in temperature is associated with a 0.45 litre reduction in average daily shower usage ($p<0.01$), alongside a marginally significant increase of 0.1 in the average number of showers per day ($p<0.1$). For air conditioning, the same temperature increase corresponds to an additional 0.21 hours of usage ($p<0.05$) and a $2.6\%$ increase in the share of residents using air conditioning on a given day ($p<0.01$). Interpreted jointly, these estimates suggest that on hotter days, residents manage thermal comfort by using more air conditioning while taking more frequent but shorter showers. Consistent with our findings, \citet{salvo2018electrical} documents that in Singapore, higher-income households increase their electricity demand, primarily through air conditioning, in response to higher temperatures.\footnote{Relatedly, \citet{zhang2022extreme} provide causal evidence that extreme weather conditions (defined as days where mean temperature exceeds $32^{\circ}$C) result in a marked increase in electricity consumption in China. This increase is attributed to the households increasing their air-conditioning usage for heat relief.} 

The key takeaway is that both shower water and air-conditioning usage are highly responsive to temperature fluctuations in our setting. This provides a useful benchmark for interpreting the effects of our interventions, both within and across domains.

%%%%%%%%%%%%%
%% Table 1 %%
%%%%%%%%%%%%%
{\singlespacing

% \begin{sidewaystable}
\begin{table}[ht]
\centering 
	\caption{Effect of Temperature on Shower and Air-Conditioning Usage} \label{tab:weather_impacts}
	\begin{threeparttable}
    \resizebox{\textwidth}{!}{% Resize table to fit within the textwidth of the page
		\begin{tabular}{l*{4}{c}}
			\toprule
			\toprule
			Dependent variable: & \multicolumn{2}{c}{Shower} & \multicolumn{2}{c}{Air Conditioning} \\ 
            \cmidrule(lr){2-3} \cmidrule(lr){4-5} 
            & \multicolumn{1}{c}{Avg. usage (litres)} & \multicolumn{1}{c}{No. of showers} & \multicolumn{1}{c}{Avg. usage (hours)} & \multicolumn{1}{c}{Fraction with usage} \\ 
            \cmidrule(lr){2-2} \cmidrule(lr){3-3} \cmidrule(lr){4-4} \cmidrule(lr){5-5} 
            & (1) & (2) & (3) & (4) \\ 
			\midrule \addlinespace 
			Average Daily Temperature ($^{\circ}$C)   &      --0.449{***}&      0.099{*} &       0.213{**}   &       0.026{***}   \\
			 &     (0.131)   &     (0.054) &     (0.092)   &     (0.008) \\ \addlinespace
			\midrule \addlinespace
			Controls & \ding{51} & \ding{51} & \ding{51} & \ding{51}  \\ \addlinespace
			\(R^{2}\) & 0.361 & 0.799 & 0.219 & 0.343 \\ \addlinespace  
			Observations & 117 & 117 & 117 & 117 \\   
			\bottomrule
		\end{tabular}
  }
		\begin{tablenotes}
			\scriptsize\vspace{0.1cm} 
            \parbox{.95\textwidth}{
			\item \emph{Notes.} This table reports OLS estimates of the relationship between average daily temperature and resource usage. Controls include daily total rainfall (mm) and an indicator for weekends. Each observation is a day during the experimental period from 10 August 2021 to 4 December 2021. Columns (1) and (2) report estimates for shower usage, measured by average water use per shower (intensive margin) and the average number of showers per day (extensive margin), respectively. Columns (3) and (4) report estimates for air-conditioning usage, measured by average daily hours (intensive margin) and the fraction of rooms with any air-conditioning usage each day (extensive margin), respectively. Robust standard errors in parentheses.
			\item * $p<0.10$, ** $p<0.05$, *** $p<0.01$
            }
		\end{tablenotes}
	\end{threeparttable}	
    % \vspace{1.0em}
\end{table}
% \end{sidewaystable}
}

\section{Results}

\subsection{Estimation Strategy}

We estimate the average treatment effects (ATE) of being assigned to real-time feedback (RTF), social comparisons (SC-S for shower usage or SC-A for air-conditioning usage), or their combination using a difference-in-differences regression model of the form:
\begin{align}
y_{it}  =  \alpha_{i}  &+ \lambda_{t} + \beta_{1} \; \text{RTF}_{it}  +  \beta_{2} \; \text{SC-S}_{it} +  \beta_{3} \; \text{SC-A}_{it} \nonumber \\
 &+  \beta_{4} \left(\text{SC-S}_{it} \times \text{RTF}_{it} \right) + \beta_{5} \left(\text{SC-A}_{it} \times \text{RTF}_{it}\right) + \epsilon_{it}, 
\end{align} 
where $y_{it}$ denotes the outcome variable, either water use per shower for device $i$ on day $t$, or total air-conditioning use for room $i$ on day $t$. $\alpha_{i}$ is the device or room fixed effect and $\lambda_{t}$ is the day fixed effect. $\text{RTF}_{it}$, $\text{SC-S}_{it}$, and $\text{SC-A}_{it}$ are indicators for device or room $i$ receiving the respective treatment on day $t$. $\epsilon_{it}$ is the random error term. Standard errors are clustered at the bathroom level, which is the unit of randomization. 

\subsection{Impact on Targeted Behavior}

Table \ref{tab:main} reports the effects of our interventions on conservation behavior in the targeted domain. Specifically, we examine how the RTF and SC-S treatments impact shower water usage, and how the SC-A treatment affects air-conditioning usage. When implemented in isolation, the RTF and SC-S treatments led to significant reductions in shower water usage by 15.9\% and 3.5\%, respectively, from baseline levels. These effects are consistent with estimates from prior studies \citep{tiefenbeck2018overcoming,agarwal2022goal, goette2021dynamics, fang2023complementarities, andor2023differences}. 
Interestingly, in column (2), we find a significant positive interaction effect of 1.708 litres $(p = 0.044)$ between the two treatments. This indicates that the marginal effect of RTF is attenuated by the addition of the SC-S intervention, and conversely, the effect of SC-S is nearly fully offset when delivered alongside RTF. As a result, the combined RTF + SC-S interventions yield an overall treatment effect that is comparable in magnitude to RTF alone.\footnote{This contrasts with the findings of \citet{andor2023differences}, who document complementarities between social comparison and real-time feedback. In their setting, social comparison was only delivered via weekly email reports to households, whereas in our setting, social comparison posters were placed directly inside the showers, where treated residents were simultaneously exposed to real-time feedback.} 

We consider two non-mutually exclusive explanations for this attenuation. One possibility is a form of attentional crowd-out: individuals may have limited capacity to attend to multiple interventions simultaneously, leading them to become less responsive to any single intervention when several are delivered at once. Another explanation relates to how residents respond to relative performance feedback, for example through moral licensing or belief-based adjustment about conservation norms. Under the combined RTF + SC-S treatment, residents are more likely to receive positive information about their relative performance than under the SC-S treatment alone. This occurs because the real-time feedback in phase 1 induces substantial conservation that improves their relative standing, which is then communicated in phase 2. If residents respond to favorable relative performance feedback by reducing conservation effort, then the additional social comparison treatment may attenuate, or even offset, the behavioral response induced by the RTF alone. We further explore this possibility in the next subsection, where we examine treatment heterogeneity by performance ranking.

By contrast, the SC-A treatment did not significantly influence daily air-conditioning usage, which aligns with findings from \citet{goette2021effects}. The null effect on air-conditioning usage is precisely estimated, and we can rule out effects as small as 12 minutes of daily usage (5.6\% of baseline) with our large sample. Importantly, the differing outcomes between the SC-S and SC-A treatments suggest that the lack of response for air-conditioning usage was not due to inattentiveness to the social information, as the subjects clearly responded to the SC-S treatment by reducing their shower usage. Our results are consistent with broader evidence suggesting that energy consumption may be more difficult to shift. In a similar university residential setting, \citet{myers2020social} document no significant effect of social comparison-based home energy reports on residents' energy consumption. Similarly, \citet{delmas2014saving} find that private real-time information and social comparisons had no direct effect on energy usage; only the addition of publicly displayed information led to significant reductions, possibly due to reputational concerns.\footnote{In their public information treatment, posters were placed in prominent high-traffic common areas (i.e., by the elevators), whereas our social comparison posters were located inside bathrooms. This limits visibility to others and renders our SC-A treatment more comparable to the private information treatment in their design.}

% Temporal dynamics of treatment effects
Additionally, to examine whether these effects vary over time, we estimate an event-study specification that interacts treatment assignment with weekly event-time indicators. Figure \ref{fig:temporal_direct} in the Appendix presents the temporal dynamics of treatment effects for the RTF and SC-S interventions on shower water usage, as well as the SC-A intervention on air-conditioning usage. For RTF, we observe a sharp and immediate reduction in shower water usage at the time of intervention, which remains stable throughout the intervention period. The effects of the SC-S intervention are noisier but do not exhibit any time trend either. For the SC-A intervention, treatment effects on air-conditioning usage remain zero throughout, reinforcing our finding of a precisely estimated null effect.  

%%%%%%%%%%%%%
%% Table 1 %%
%%%%%%%%%%%%%
{\singlespacing

% \begin{sidewaystable}
\begin{table}[ht]
\centering 
	\caption{Effects of Real-Time Feedback and Social Comparisons} \label{tab:main}
	\begin{threeparttable}
    \resizebox{\textwidth}{!}{% Resize table to fit within the textwidth of the page
		% \begin{tabular}{l*{4}{c}}
        \begin{tabular}{l*{4}{>{\centering\arraybackslash}p{2.2cm}}}
			\toprule
			\toprule
			Dependent variable: & \multicolumn{2}{c}{Water use per shower (litres)} & \multicolumn{2}{c}{Aircon use per day (hours)} \\ 
            \cmidrule(lr){2-3} \cmidrule(lr){4-5} 
            & (1) & (2) & (3) & (4) \\ 
			\midrule \addlinespace 
			Real-Time Feedback (RTF)    &      --5.490{***}&      --5.959{***} &       0.016   &       0.076   \\
			 &     (0.478)   &     (0.501) &     (0.110)   &     (0.110)   \\ \addlinespace
			Social Comparison for Shower (SC-S) &      --1.211{**}  &      --2.014{***} &      --0.014   &       0.102   \\
            &     (0.597)   &     (0.696) &     (0.111)   &     (0.137)   \\ \addlinespace
            Social Comparison for Aircon (SC-A) &      --0.340   &      --1.159  &      --0.086   &      --0.005   \\
			&     (0.679)   &     (0.929)  &     (0.142)   &     (0.168)   \\ \addlinespace
            SC-S $\times$ RTF &               &       1.708{**} &               &      --0.231   \\
            &               &     (0.743) &      &     (0.174)   \\ \addlinespace
            SC-A $\times$ RTF &               &       1.592 &               &      --0.159   \\
            &    &     (0.994) &               &     (0.241)   \\ \addlinespace
            \addlinespace
            Baseline Mean & 34.565 & 34.565 & 3.567 & 3.567 \\
                 & (27.283)   &   (27.283) &  (5.059)  &  (5.059)  \\ 
        % Constant & 35.686{***} & 35.855{***} &       3.805{***}&       3.788{***} \\
        %      &     (0.187)   &     (0.200) &  (0.038)   &     (0.036)  \\ 
             \addlinespace
			% \midrule \addlinespace
   %          $p$-values of interest: \\  \addlinespace
			% Level differences in TEs: $\beta_{18L} = \beta_{28L}$ &  0.919 & 0.070 &  0.348 & 0.014 \\  \addlinespace
   %          Divergence in TEs: $(\beta_{18L2} - \beta_{18L1}) = (\beta_{28L2} - \beta_{28L1})$ &  \multicolumn{2}{c}{0.006} &  \multicolumn{2}{c}{0.006} & \multicolumn{2}{c}{0.097} \\  \addlinespace
			% Change of TE over time: $\beta_{18L1} = \beta_{18L2}$ &  \multicolumn{2}{c}{0.059} &  \multicolumn{2}{c}{0.042} &  \multicolumn{2}{c}{0.349}  \\ \addlinespace
			% Change of TE over time: $\beta_{28L1} = \beta_{28L2}$ &  \multicolumn{2}{c}{0.455} &  \multicolumn{2}{c}{0.436} &  \multicolumn{2}{c}{0.717}  \\ \addlinespace
			% $F$-test & \multicolumn{2}{c}{0.000} & \multicolumn{2}{c}{0.000} & \multicolumn{2}{c}{0.000} \\  \addlinespace    
			\midrule \addlinespace
			Device/Room FEs & \ding{51} & \ding{51} & \ding{51} & \ding{51}   \\ \addlinespace
			Date FEs & \ding{51} & \ding{51} & \ding{51} & \ding{51}  \\ \addlinespace
			\(R^{2}\) & 0.227 & 0.227 & 0.473 & 0.473 \\ \addlinespace  
			Observations & 153,882 & 153,882 & 147,376 & 147,376 \\   
			\bottomrule
		\end{tabular}
  }
		\begin{tablenotes}
			\scriptsize\vspace{0.1cm} 
            \parbox{.95\textwidth}{
			\item \emph{Notes.} This table reports OLS estimates of the treatment effects of the RTF, SC-S, and SC-A interventions. All specifications include device/room and day fixed effects. Columns (1) and (2) report estimates for shower water usage, while columns (3) and (4) report estimates for daily air-conditioning usage. Standard errors clustered at the bathroom level in parentheses.
			\item * $p<0.10$, ** $p<0.05$, *** $p<0.01$
            }
		\end{tablenotes}
	\end{threeparttable}	
    % \vspace{1.0em}
\end{table}
% \end{sidewaystable}
}

\subsection{Impact on Non-Targeted Behavior}

We next explore potential spillover effects of our interventions by assessing their impact on consumption behavior in the non-targeted domain. Specifically, we examine the effect of the RTF and SC-S treatments on air-conditioning usage, and the effect of the SC-A treatment on shower water usage. Returning to Table \ref{tab:main}, our findings reveal that while the RTF intervention leads to significant reductions in shower water usage (i.e., a strong first-stage effect), it does not induce any spillover effects on air-conditioning usage. Notably, our large sample allows us to reject an effect size as small as 12 minutes of daily air-conditioning usage. Therefore, we summarize our results as follows:

\begin{customres}{1:}{\label{res: 1}}
\textbf{Null cross-domain spillovers.} The RTF intervention reduces shower water usage (targeted behavior), but has no significant effect on air-conditioning usage (non-targeted behavior), on average.
\end{customres}

% Add some speculative discussion about why RTF clearly draws attention to water usage but does not generate spillovers (in contrast to other studies cited)
These findings are not consistent with attentional spillovers between water and energy consumption in our setting, nor with a thermal comfort substitution channel, whereby reductions in shower water use lead residents to compensate through increased air-conditioning usage for heat relief. One possible explanation for the absence of attentional spillovers is that the RTF intervention draws attention to water usage only at a very specific moment, during the act of showering. Because the feedback is tightly coupled with the immediate context in which the resource is consumed, it may not be salient enough outside the shower to redirect attention towards or away from other forms of consumption, such as air conditioning.\footnote{This contrasts with other studies that documented attentional spillovers using reminder messages, which are more broadly framed and delivered outside the immediate context of the target behavior, such as prompts to log meals or make credit card payments \citep{trachtmanndoes, medina2021side}.} Thus, the two resource domains do not act as attentional substitutes or complements in our setting.

A similar pattern is observed with the SC-S treatment; it has no significant effect on air-conditioning usage, despite a modest but significant effect on shower water usage. For the SC-A treatment, the lack of a direct effect on air-conditioning usage appears to preclude any indirect spillover effects on shower water usage. 

\subsection{Treatment Heterogeneity by Performance Ranking}

We next assess whether responses vary with the relative performance information conveyed through the social comparison interventions. Recall that in both the SC-S and SC-A treatments, residents receive a ranking based on their bathroom's average shower or air-conditioning usage. Table \ref{tab:main2} presents the ATEs of our interventions from a specification augmented with interactions between treatment assignment and percentile rankings. Specifically, $\text{PercentileRank}_{i}^{S}$ and $\text{PercentileRank}_{i}^{A}$ are fixed bathroom-level ranking measures based on shower and air-conditioning usage, respectively, in the two-week window immediately preceding phase 2. They therefore represent each bathroom's initial ranking prior to the social comparison interventions. We normalize these percentile rankings to a scale from $-50$ to 50, where a value of 0 corresponds to the median-ranked bathroom within the same residential college and room type, and higher values indicate more favorable rankings (i.e., lower usage). The main treatment effects are thus interpreted as the impact of the respective interventions for bathrooms at the median ranking. We also include interactions between the percentile measures and the phase 2 indicator to account for possible dynamics correlated with percentiles (e.g., regression to the mean). The percentile rankings of the control group identify these coefficients. Appendix B.1 provides further details on the construction of the percentile measures.

In column (1), we observe a robust and significant interaction effect for the SC-S intervention by performance ranking, with a point estimate of 0.061 litres ($p < 0.01$). The positive coefficient indicates that residents using bathrooms with higher rankings, meaning those with lower pre-phase-2 shower water usage, responded less to the SC-S intervention. The treatment effect reverses sign around the 70th percentile. To illustrate, residents using bathrooms at the median ranking reduced shower water usage by $-$1.326 litres ($p = 0.029$). Residents using bathrooms at the 10th percentile exhibited a more pronounced conservation effect, reducing usage by $-$3.779 litres, while those at the 90th percentile increased usage by 1.135 litres---a highly significant difference in treatment effects ($p < 0.01$). 

This pattern indicates that responses to social comparisons vary meaningfully across the performance distribution in the showering domain. One possible interpretation is moral licensing, under which residents using bathrooms that were ranked favorably relative to their peers reduce conservation effort, or even increase usage, in the subsequent period.\footnote{Aligning with our results, \citet{agarwal2022water} find heterogeneous effects of nationwide peer comparisons on residential water consumption using a quasi-experimental design: households with below-median baseline consumption increased their water usage post-treatment, while those above the median decreased theirs.} However, the same pattern may also reflect norm adherence or conformity, whereby residents using higher-ranked bathrooms adjust their usage toward the perceived social norm.

In contrast, column (2) reveals no significant interaction between the SC-A treatment and the air-conditioning percentile ranking, providing no evidence of within-domain heterogeneity in the air-conditioning domain. We also do not observe cross-domain treatment heterogeneity: the coefficient on $\text{SC-A}_{it}$ $\times$ $\text{PercentileRank}_{i}^{A}$ in column (1) for shower water use, and analogously the coefficient on $\text{SC-S}_{it}$ $\times$ $\text{PercentileRank}_{i}^{S}$ in column (2) for air-conditioning use are both statistically insignificant. We summarize these findings as follows:

\begin{customres}{2:}{\label{res: 2a}}
\textbf{Within-domain heterogeneity by performance ranking observed in the showering domain but not the air-conditioning domain.} Within the showering domain, the treatment effect of SC-S decreases with performance ranking in shower water usage. No significant heterogeneity by performance ranking is observed for the SC-A treatment in air-conditioning usage.
\end{customres}

\begin{customres}{3:}{\label{res: 2b}}
\textbf{No cross-domain heterogeneity by performance ranking.} The treatment effect of SC-A on shower water use and the treatment effect of SC-S on air-conditioning use do not vary significantly with performance ranking.
\end{customres}

%%%%%%%%%%%%%
%% Table 3 %%
%%%%%%%%%%%%%
{\singlespacing

% \begin{sidewaystable}
\begin{table}[H]
\centering 
	\caption{Heterogeneous Effects by Performance Ranking} \label{tab:main2}
	\begin{threeparttable}
    \resizebox{\textwidth}{!}{% Resize table to fit within the textwidth of the page
		\begin{tabular}{l*{2}{c}}
			\toprule
			\toprule
			Dependent variable: & \multicolumn{1}{c}{Water use per shower (litres)} & \multicolumn{1}{c}{Aircon use per day (hours)} \\ 
            % \addlinespace
            \cmidrule(lr){2-2} \cmidrule(lr){3-3} 
            % \addlinespace
            % & \multicolumn{2}{c}{Full Sample} & \multicolumn{2}{c}{Above Median} & \multicolumn{2}{c}{Below Median}  \\ 
            & (1) & (2)  \\ 
			\midrule \addlinespace 
			Real-Time Feedback (RTF)  & --5.209{***} &      --0.060   \\
			 &  (0.547) &     (0.107) \\ \addlinespace
    Social Comparison for Shower (SC-S) &--1.322{**} &      --0.058   \\
            &     (0.600)  &     (0.104)  \\ \addlinespace
			Social Comparison for Aircon (SC-A) &--0.321 & --0.135  \\
			 &   (0.694) &  (0.127)  \\ \addlinespace
            RTF $\times$ $\text{PercentileRank}^{S}$  & --0.034{*} & 0.005   \\
			 &  (0.021) &    (0.003) \\ \addlinespace
            SC-S $\times$ $\text{PercentileRank}^{S}$  &  0.061{***} & --0.007  \\
             &     (0.023)  &     (0.004) \\ \addlinespace
             SC-A $\times$ $\text{PercentileRank}^{A}$ &   0.020 &       0.005  \\
             &     (0.019) &     (0.005)   \\ \addlinespace
            Phase 2 $\times$ $\text{PercentileRank}^{S}$ &  --0.033{*}  & 0.004 \\
               &     (0.017)    &     (0.003)  \\ \addlinespace
               Phase 2 $\times$ $\text{PercentileRank}^{A}$ &      --0.012  &       0.001  \\
               &     (0.009)   &     (0.003) \\ \addlinespace
               \addlinespace
            Baseline Mean & 34.565  & 3.567 \\
                 & (27.283)  &  (5.059)  \\ 
        % Constant&      35.522{***} &     3.834{***}\\
        %      &     (0.204) &     (0.036) \\ 
             \addlinespace
			% \midrule \addlinespace
   %          $p$-values of interest: \\  \addlinespace
			% Level differences in TEs: $\beta_{18L} = \beta_{28L}$ &  0.919 & 0.070 &  0.348 & 0.014 \\  \addlinespace
   %          Divergence in TEs: $(\beta_{18L2} - \beta_{18L1}) = (\beta_{28L2} - \beta_{28L1})$ &  \multicolumn{2}{c}{0.006} &  \multicolumn{2}{c}{0.006} & \multicolumn{2}{c}{0.097} \\  \addlinespace
			% Change of TE over time: $\beta_{18L1} = \beta_{18L2}$ &  \multicolumn{2}{c}{0.059} &  \multicolumn{2}{c}{0.042} &  \multicolumn{2}{c}{0.349}  \\ \addlinespace
			% Change of TE over time: $\beta_{28L1} = \beta_{28L2}$ &  \multicolumn{2}{c}{0.455} &  \multicolumn{2}{c}{0.436} &  \multicolumn{2}{c}{0.717}  \\ \addlinespace
			% $F$-test & \multicolumn{2}{c}{0.000} & \multicolumn{2}{c}{0.000} & \multicolumn{2}{c}{0.000} \\  \addlinespace    
			\midrule \addlinespace
			Device/Room FEs &  \ding{51} &  \ding{51}   \\ \addlinespace
			Date FEs  & \ding{51} &  \ding{51} \\ \addlinespace
			\(R^{2}\) & 0.220 & 0.473 \\ \addlinespace  
			Observations & 142,860 & 144,702 \\   
			\bottomrule
		\end{tabular}
  }
		\begin{tablenotes}
			\scriptsize\vspace{0.1cm} 
            \parbox{.95\textwidth}{
			\item \emph{Notes.} This table reports OLS estimates of the treatment effects of the RTF, SC-S, and SC-A interventions by performance ranking. All specifications include device/room and day fixed effects. Column (1) reports estimates for shower water usage, while column (2) reports  estimates for daily air-conditioning usage. $\text{PercentileRank}^{S}$ and $\text{PercentileRank}^{A}$ are fixed bathroom-level percentile rankings based on shower and air-conditioning usage, respectively, in the two-week window immediately preceding phase 2. Both measures are normalized to range from $-50$ to $50$, with higher values indicating more favorable rankings (i.e., lower usage). Standard errors clustered at the bathroom level in parentheses.
   \item * $p<0.10$, ** $p<0.05$, *** $p<0.01$
   }
		\end{tablenotes}
	\end{threeparttable}	
    % \vspace{1.0em}
\end{table}
% \end{sidewaystable}
}

Finally, we examine whether the absence of spillover effects holds consistently over time by estimating an event-study specification for the non-targeted outcomes. In Appendix Figure \ref{fig:temporal_spillover}, the treatment effects of the RTF and SC-S interventions on air-conditioning usage, as well as the SC-A intervention on shower usage, remain flat and statistically indistinguishable from zero throughout the intervention period. These results reinforce our central finding: we find no evidence of spillover across domains, with precisely estimated null effects both on average and over time.

\subsection{Unpacking the Null Effects on Air-Conditioning Usage}

Across all analyses, we observe precise null effects on air-conditioning usage, both for the direct effects of the SC-A treatment and for spillover effects from the RTF and SC-S treatments. One possible explanation relates to the institutional features of our setting. Air conditioning is priced on a pay-as-you-use basis, and residents must actively purchase credits, making usage costs salient. This pricing structure may encourage residents to closely monitor air-conditioning usage and weigh it against their thermal comfort preferences and budget constraints. While feedback interventions can be effective in settings where individuals misperceive their energy consumption \citep{gerster2025disaggregate}, such informational frictions may be less relevant in our setting, where residents regularly monitor their prepaid balances. As a result, there may be limited scope for additional behavioral adjustment in response to our interventions. 

We examine this interpretation by testing for treatment heterogeneity along three margins where adjustment might be more likely: baseline air-conditioning usage, daily temperature, and budget constraints proxied by room type. In particular, high baseline users may have more scope to reduce consumption, hotter days increase air-conditioning demand and may change adjustment margins, and residents staying in costlier apartment suites may face weaker budget constraints.

Across all three dimensions, we find no evidence of meaningful treatment heterogeneity, and the effects on air-conditioning usage are consistently null. Specifically, the SC-A treatment effect does not vary significantly across baseline usage quartiles, temperature bins, or room types; nor do we find heterogeneous spillover effects from the RTF or SC-S treatments on air-conditioning usage. These results are reported in Appendix A.5. 

Taken together, the null effects on air-conditioning usage are unlikely to mask offsetting responses. They are consistent with the interpretation that residents may already be closely monitoring their air-conditioning usage, leaving limited scope for further response to our interventions.

\section{Discussion}

A key challenge in identifying behavioral spillovers in resource consumption stems from the confounding influence of many appliances (e.g., dishwasher, washing machine) that simultaneously utilize both water and energy, referred to as ``mechanical complementarities.'' \cite{jessoe2021spillovers} take an important first step towards addressing this challenge by bounding the extent of mechanical complementarities with simulated electricity usage data to identify spillover effects, and they document reductions in summertime energy use from their social norms water-saving intervention. While their innovative approach necessarily involves assumptions about appliance use and ownership patterns, our study builds on their work by directly measuring water usage in the showers and energy usage from air conditioning, rather than relying on aggregate consumption data. Therefore, we can definitively rule out mechanical complementarities in our setting, providing a clean test of spillover effects that arise solely from shifts in behavior.

We design our field experiment to test for bidirectional spillovers between shower water usage and air-conditioning energy usage. We first consider the RTF intervention. The potential for attentional spillovers in resource conservation is particularly relevant in today's context, as consumers increasingly receive real-time feedback on a wide range of behaviors through Internet-of-Things devices. Such feedback could direct attention towards or away from related consumption decisions in other domains, especially under limited attention. With the RTF treatment, we document significant conservation effects, achieving a 15.9\% reduction in shower water use relative to baseline levels. However, we find no evidence of spillovers to the non-targeted domain, with tightly estimated null effects on air-conditioning usage. Notably, our findings highlight a key distinction between real-time feedback and reminder-based nudges. While reminders have been shown to generate negative spillover effects on other actions \citep{koch2024spillover, medina2021side}, our RTF intervention does not produce such effects on air-conditioning usage. This suggests that real-time feedback may be an effective way to influence targeted behavior without generating unintended negative responses in other domains. We view this result as a novel and important contribution to the literature.

Turning to social comparisons, the SC-S treatment generates a more modest but significant reduction in shower water usage, while the SC-A treatment has no detectable effect on air-conditioning usage. We use the performance ranking information embedded in the social comparison interventions to examine whether responses vary with relative performance within and across domains. Within the showering domain, we find significant heterogeneity in treatment effects by performance ranking: residents receiving more favorable rankings exhibit weaker conservation responses.\footnote{This result parallels the classic ``boomerang effect'' documented in prior work (e.g., \citealp{schultz2007constructive, byrne2018tell, papineau2022experimental}), where high-performing individuals respond less to conservation nudges or even increase consumption.} However, we do not observe the same treatment heterogeneity within the air-conditioning domain, nor do we find evidence of cross-domain heterogeneity. While our design does not precisely identify the mechanism underlying the treatment heterogeneity in the showering domain, the pattern is directionally consistent with moral licensing and may also reflect norm adherence, conformity, or conditional cooperation \citep{cialdini1990focus, nolan2008normative, hage2009norms, allcott2011social, frey_social_2004}. Future work could build on our design by experimentally varying the salience of normative benchmarks or decoupling performance ranking feedback from norm-based messaging to better distinguish among these mechanisms.

Overall, showering behavior responds strongly to both the RTF and SC-S treatments, while air-conditioning usage appears more difficult to influence, despite being highly responsive to local weather conditions. We find no significant direct or spillover effects on air-conditioning use, even though residents were repeatedly primed to reduce their consumption throughout the intervention period.

To situate our findings within the broader literature on behavioral spillovers in resource conservation, the mixed evidence to date underscores that spillover effects are highly context-dependent, shaped by the specific behaviors targeted, the types of interventions, and the decision environment \citep{tiefenbeck2013better, jessoe2021spillovers, carlsson2021behavioral, goetz2024one}. We note that unlike \citet{goetz2024one}, who observe positive spillovers to room heating energy consumption, we find no corresponding spillovers from both our water-saving interventions to air-conditioning usage. Consistent with the discussion in Section 3.5, one possible explanation relates to how these end uses are operated in the two settings. In \citet{goetz2024one}, households could freely adjust thermostat settings. Indeed, the authors note that the large reductions in room heating energy in their context are consistent with households lowering their thermostat once at the start of the heating period and not re-adjusting it thereafter. By contrast, our residents cannot change thermostat temperatures and need to actively choose the durations of cooling airflow whenever they wish to lower the room temperature. More speculatively, while inattention may have helped to generate a positive spillover effect in \cite{goetz2024one}'s setting by allowing an initial adjustment of the thermostat to persist, it may have worked against positive spillovers in our setting where residents must make an active choice at every juncture.

Both \citet{tiefenbeck2013better} and \cite{jessoe2021spillovers} find spillovers from water-saving interventions during the summer, a period that coincides with peak electricity consumption. These high-demand periods may offer greater scope for user adjustments in resource usage. In contrast, we examine spillover effects without seasonal peaks in resource demand in our context, since the temperature in Singapore remains stable year-round. 

Additionally, a distinct feature of our study is that we are able to isolate the treatment effects of our interventions attributable purely to behavior change, as residents do not have the ability or incentive to invest in water- or energy-efficient capital stock. By contrast, in household residential settings where such investments are feasible, similar interventions may interact with technology adoption (e.g., \citealp{brandon2017effects}), potentially generating spillovers over time. An important avenue for future research would be to better understand the longer-run spillover impacts of these interventions.

Finally, identifying spillovers in resource conservation is important for conducting a comprehensive welfare analysis of these widely-used behavioral interventions \citep{allcott2019welfare, jessoe2021spillovers}. Without accounting for spillover effects, assessments of cost-effectiveness may be significantly skewed. While we do not detect cross-domain spillovers in our setting, a full welfare evaluation must also consider potential comfort effects and unintended psychological costs. For the RTF treatment, reductions in shower usage could lower residents’ comfort levels and thereby reduce welfare. At the same time, if the feedback corrects misperception about resource use intensity and leads residents to adjust usage accordingly, this would generate consumer surplus gains (e.g., \citealp{gerster2025disaggregate}). The social comparison treatments raise a different concern: residents who receive low rankings may experience negative image utility, such as shame or social pressure. As discussed in \citet{butera2022measuring}, the net welfare effect depends on how individuals value public recognition and how image costs vary across the performance distribution. The relevant policy question is therefore whether social comparisons are more or less socially efficient than alternative policy instruments, such as direct financial incentives.

Overall, our study contributes to a better understanding of how behavioral interventions in one resource domain may influence behavior in another, helping inform the design of more holistic conservation policies. Future work could explore other interventions that more effectively influence air-conditioning usage and test for spillover effects across additional resource domains.

%--------------------------------------------------------------------
% BIBLIOGRAPHY
%--------------------------------------------------------------------
\newpage
\singlespacing
\bibliographystyle{apalike}
\bibliography{Ref}

%--------------------------------------------------------------------
%           APPENDIX          
%--------------------------------------------------------------------
\pagebreak 
\appendix
\setcounter{page}{1}

% Begin collecting a separate appendix-only contents list
\startcontents[appendix]

\setcounter{table}{0} 
\setcounter{figure}{0}
\renewcommand{\thetable}{A\arabic{table}}
\renewcommand{\thefigure}{A\arabic{figure}}

\begin{center}
    {\LARGE{\textbf{Online Appendix}}}\\[1.5em]
    {\Large Testing for Spillovers in Resource Conservation:}\\
    {\Large Evidence from a Natural Field Experiment}\\[1em]
    {\large Lorenz Goette \qquad Zhi Hao Lim}
\end{center}

\vspace{1.5em}

\begingroup
\hypersetup{hidelinks}
\printcontents[appendix]{}{1}[2]{}
\endgroup

\clearpage

\section{Supplementary Figures and Tables}

\subsection{Experimental Setting and Intervention Materials}

\begin{figure}[H] 
\center
\caption{Layout of the Residential College} \label{fig:floor_plan}
	\includegraphics[width=\textwidth]{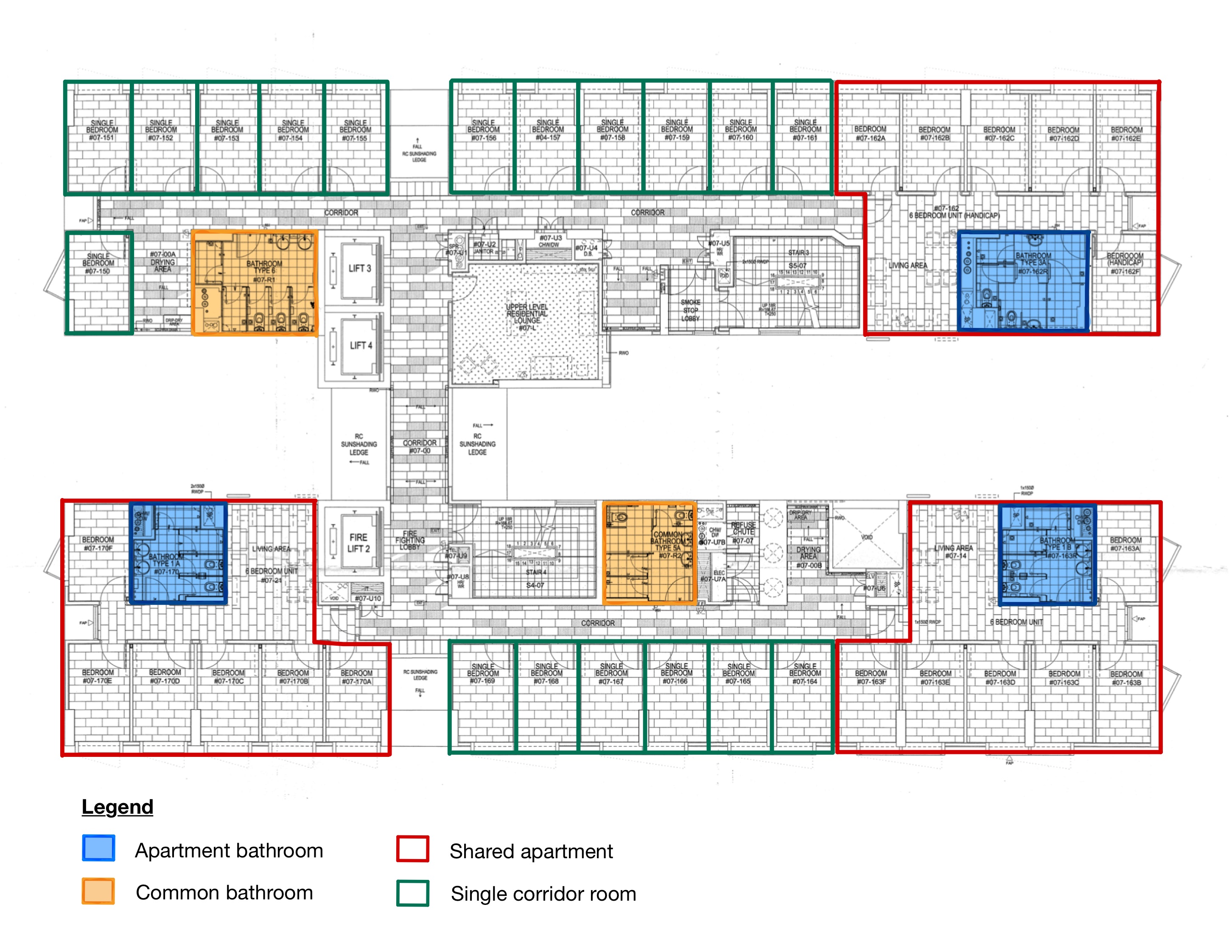} \\[2mm]
	\begin{minipage}{15cm}
		\footnotesize \emph{Notes.} This figure displays a representative floor plan of the residential colleges, illustrating that each floor contains two types of bathrooms: the common bathroom, marked in orange, and the apartment (suite) bathroom, marked in blue. The unit of randomization is at the bathroom level. See \url{https://uci.nus.edu.sg/ohs/future-residents/undergraduates/utown/room-types/} for further details. 
	\end{minipage}
\end{figure}

\newpage
\begin{figure}[H]
	\centering
    \caption{Sample Poster by Experimental Condition}
    \vspace{0.5em}

    % Control
	\begin{subfigure}{0.45\textwidth}
    \caption{Control}
		\includegraphics[width=\textwidth]{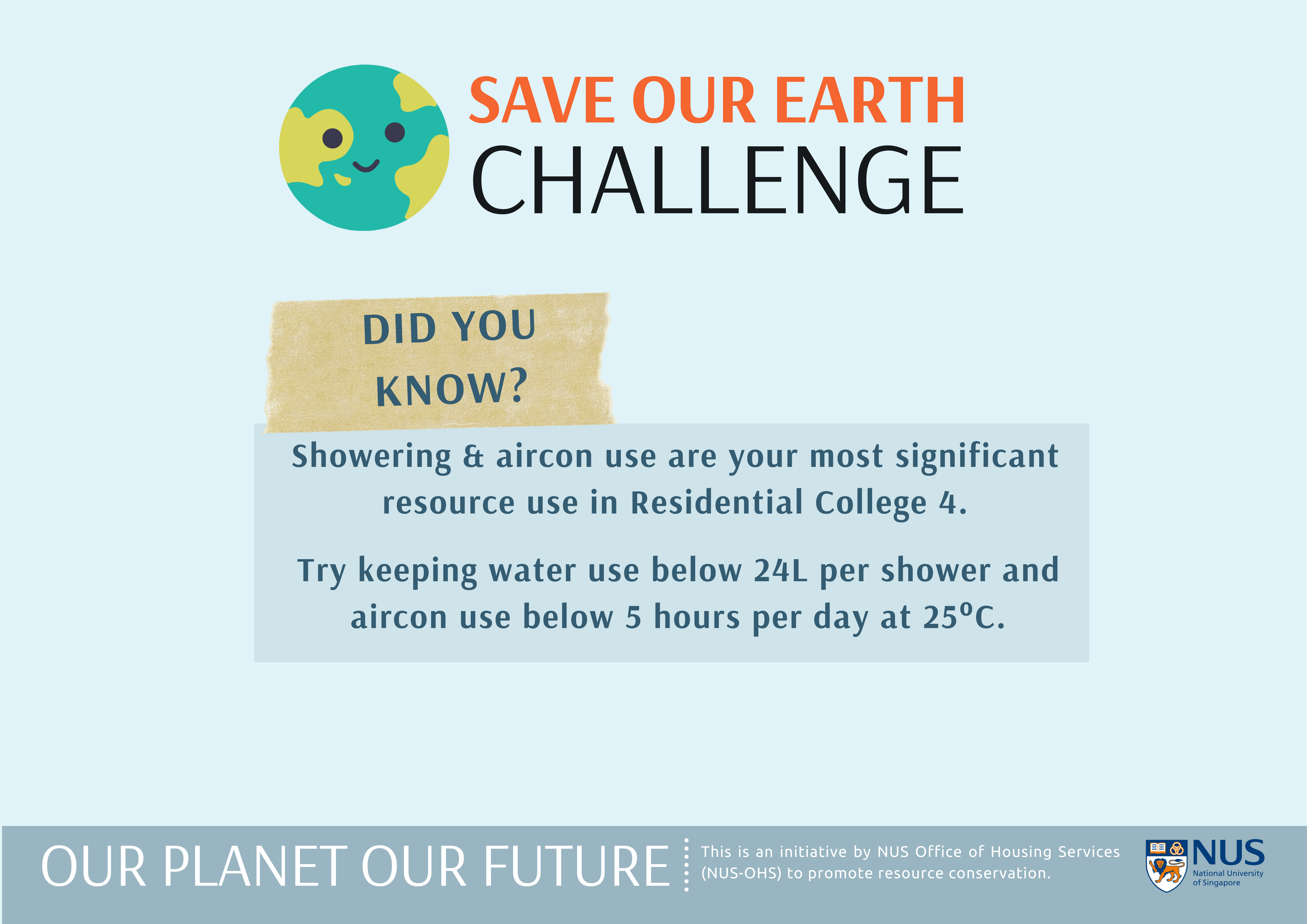} 
  % \label{fig:condition1}
  % \vspace{0.5em}
	\end{subfigure}
	\begin{subfigure}{0.45\textwidth}
    \caption{SC-S}
		\includegraphics[width=\textwidth]{figures/Treatment_3-min.png} 
  % \label{fig:condition2}
  % \vspace{0.5em}
	\end{subfigure}
	\begin{minipage}{15cm}
		\vspace{.5em}
		\footnotesize \emph{} 
	\end{minipage}

	% SC-A
	\begin{subfigure}{0.45\textwidth}
    \caption{SC-A} 	
		\includegraphics[width=\textwidth]{figures/Treatment_2-min.png} 
  % \label{fig:condition3}
  % \vspace{0.5em}
	\end{subfigure}
    % RTF
	\begin{subfigure}{0.45\textwidth}
    \caption{RTF}
		\includegraphics[width=\textwidth]{figures/Treatment_4-min.png} 
  % \label{fig:condition4}
  % \vspace{0.5em}
	\end{subfigure}
	\begin{minipage}{15cm}
		\vspace{.5em}
		\footnotesize \emph{} 
	\end{minipage}
	
    % RTF + SC-S
	\begin{subfigure}{0.45\textwidth}
    \caption{RTF + SC-S}
		\includegraphics[width=\textwidth]{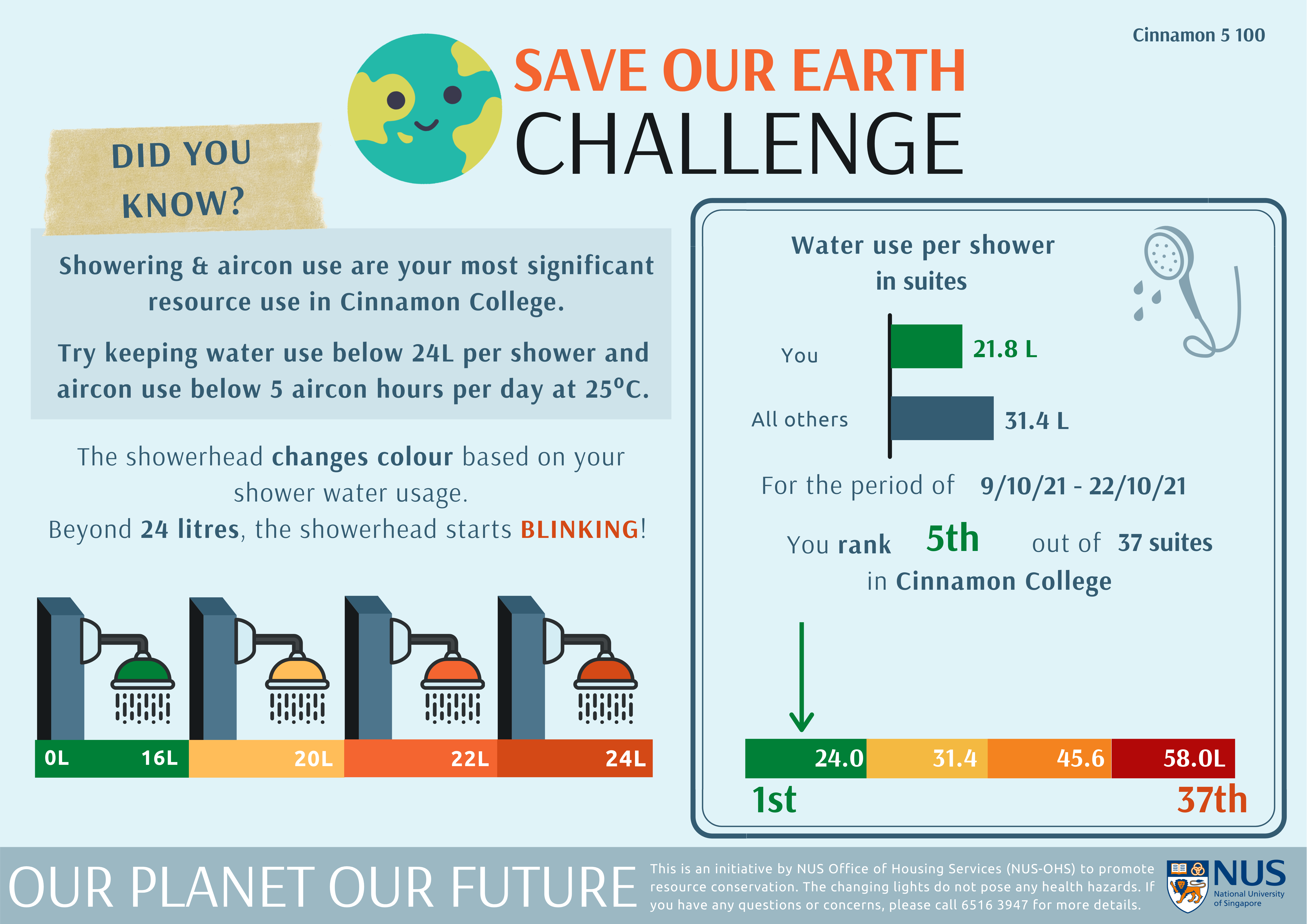} 
  % \label{fig:condition5}
  % \vspace{0.5em}
	\end{subfigure}
    % RTF + SC-A
	\begin{subfigure}{0.45\textwidth}
    \caption{RTF + SC-A}
		\includegraphics[width=\textwidth]{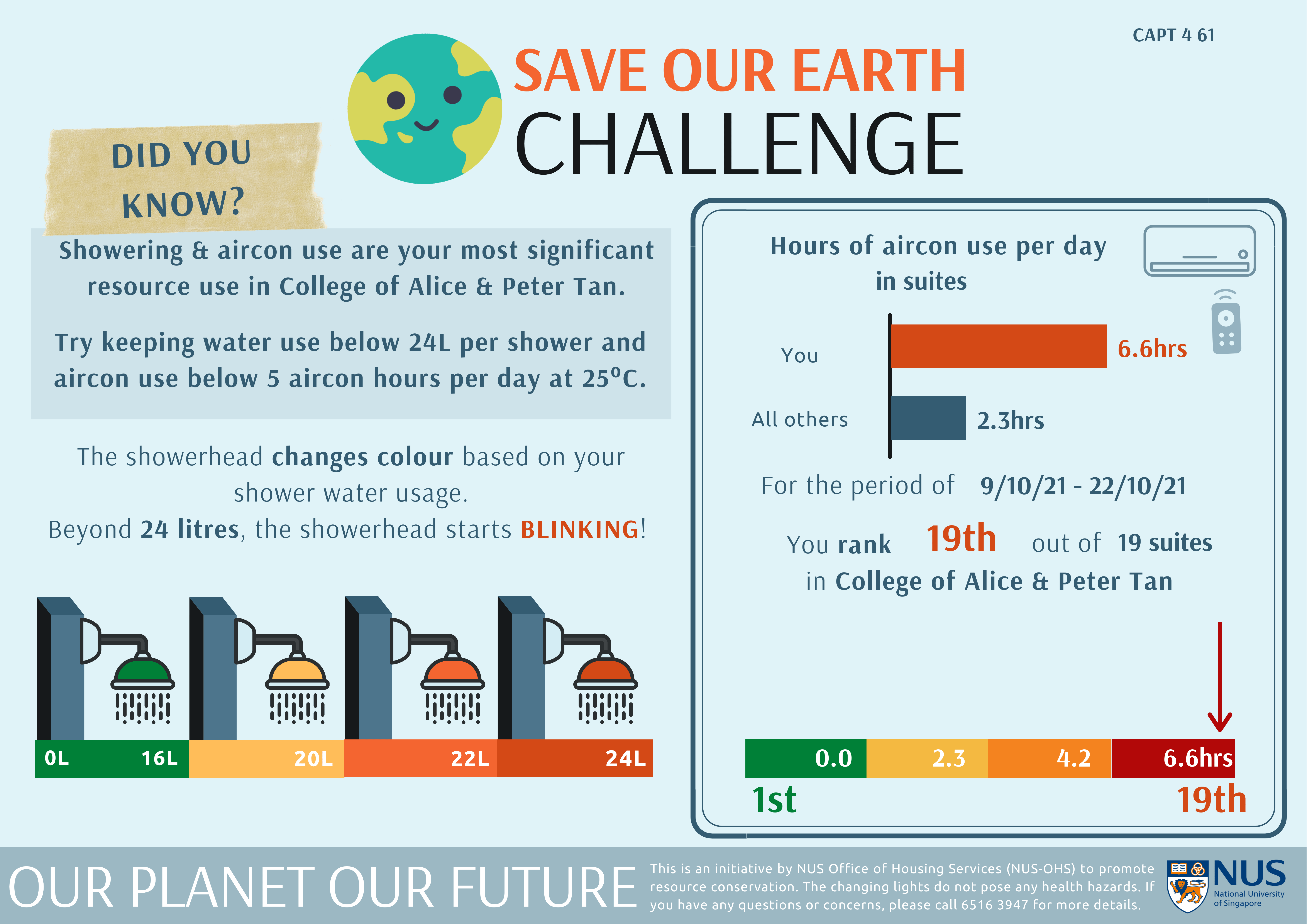} 
  % \label{fig:condition6}
  % \vspace{0.5em}
	\end{subfigure}
 
	\begin{minipage}{15cm}
		\vspace{2.0em}
		\footnotesize \emph{Notes.} A specific type of poster was displayed in each shower facility based on the assigned experimental condition. Since the interventions were rolled out in two phases, the posters from phase 1 (either Control or RTF) were replaced with new, corresponding posters (featuring social comparison of either shower or air-conditioning usage) upon transitioning to phase 2 of the experiment.
	\end{minipage}
	
\label{fig:sample_posters} 
\end{figure}

\begin{figure}[H] 
\caption{Real-Time Feedback (RTF) Intervention} 
\label{fig: rtf_treatment} 
% \vspace{1.0em}
	\begin{center}
	\begin{subfigure}{0.24\textwidth}
		\centering
        \caption{Below 14L}
		\includegraphics[angle=360,width=.95\textwidth]{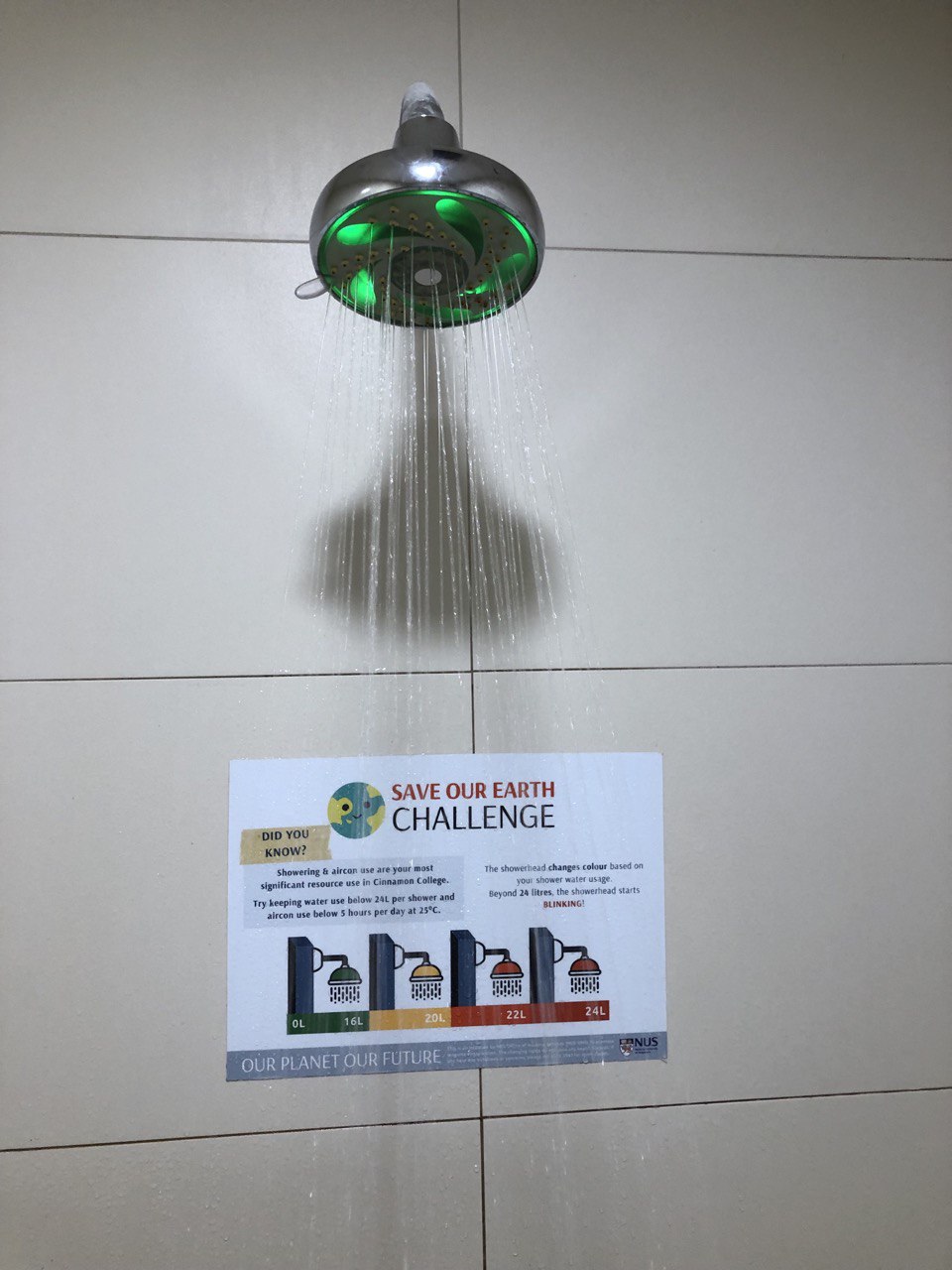}
	\end{subfigure}
	\begin{subfigure}{0.24\textwidth}
		\centering
        \caption{14L to 20L}
		\includegraphics[angle=360,width=.95\textwidth]{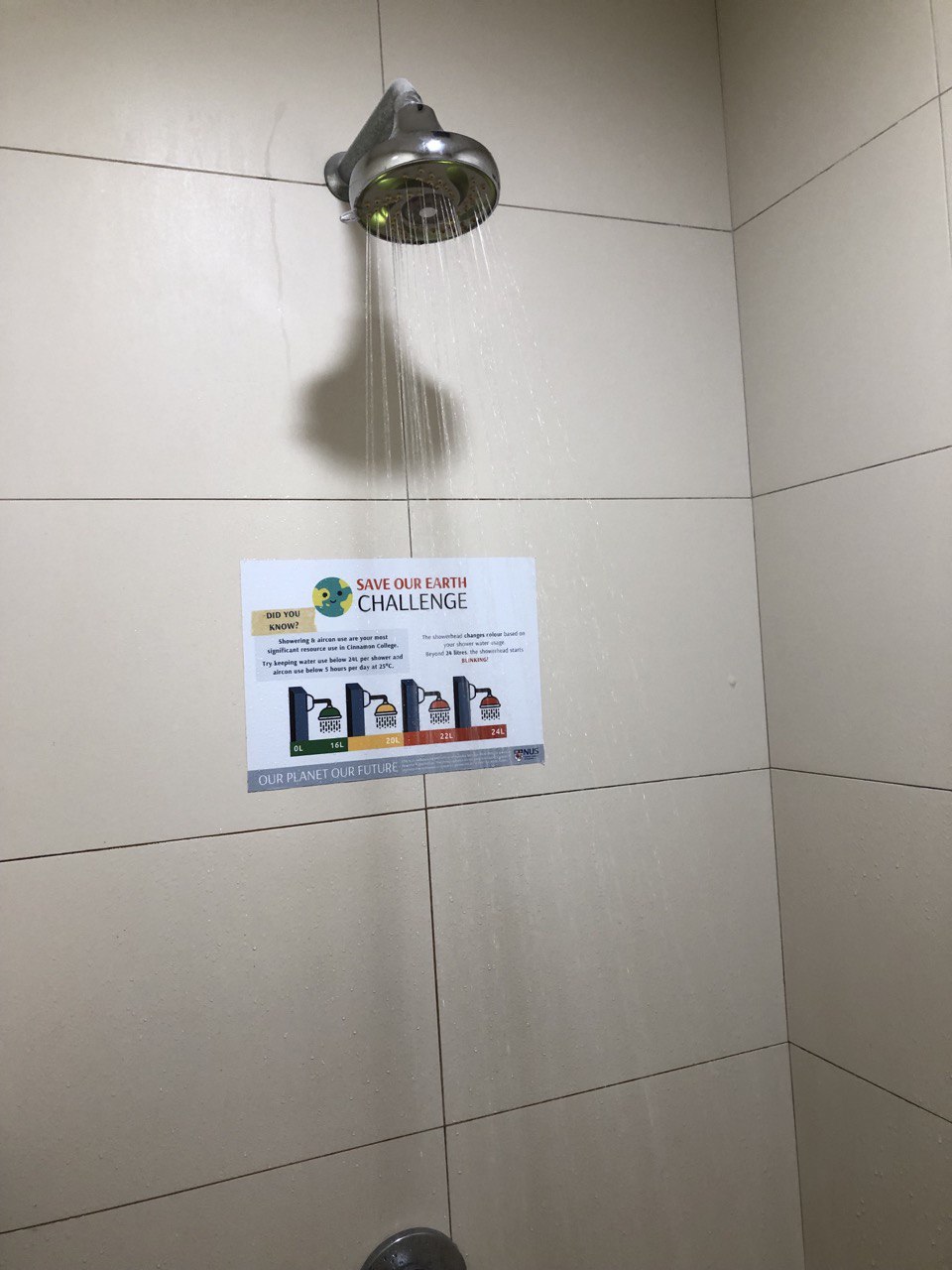}
	\end{subfigure} 
	\begin{subfigure}{0.24\textwidth}
		\centering
        \caption{20L to 22L}
		\includegraphics[angle=360,width=.95\textwidth]{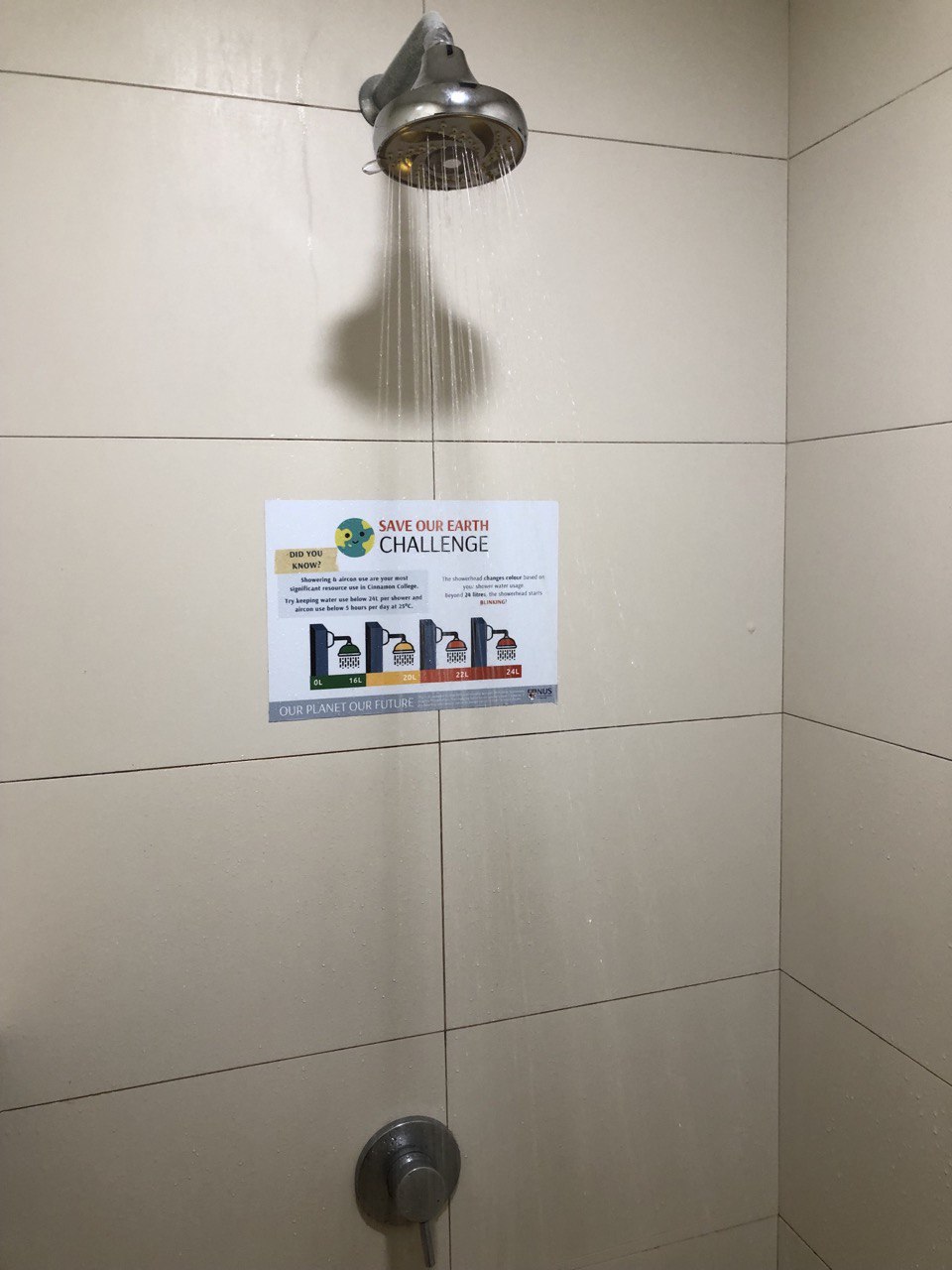}
	\end{subfigure}
	\begin{subfigure}{0.24\textwidth}
		\centering
        \caption{22L to 24L}
		\includegraphics[angle=360,width=.95\textwidth]{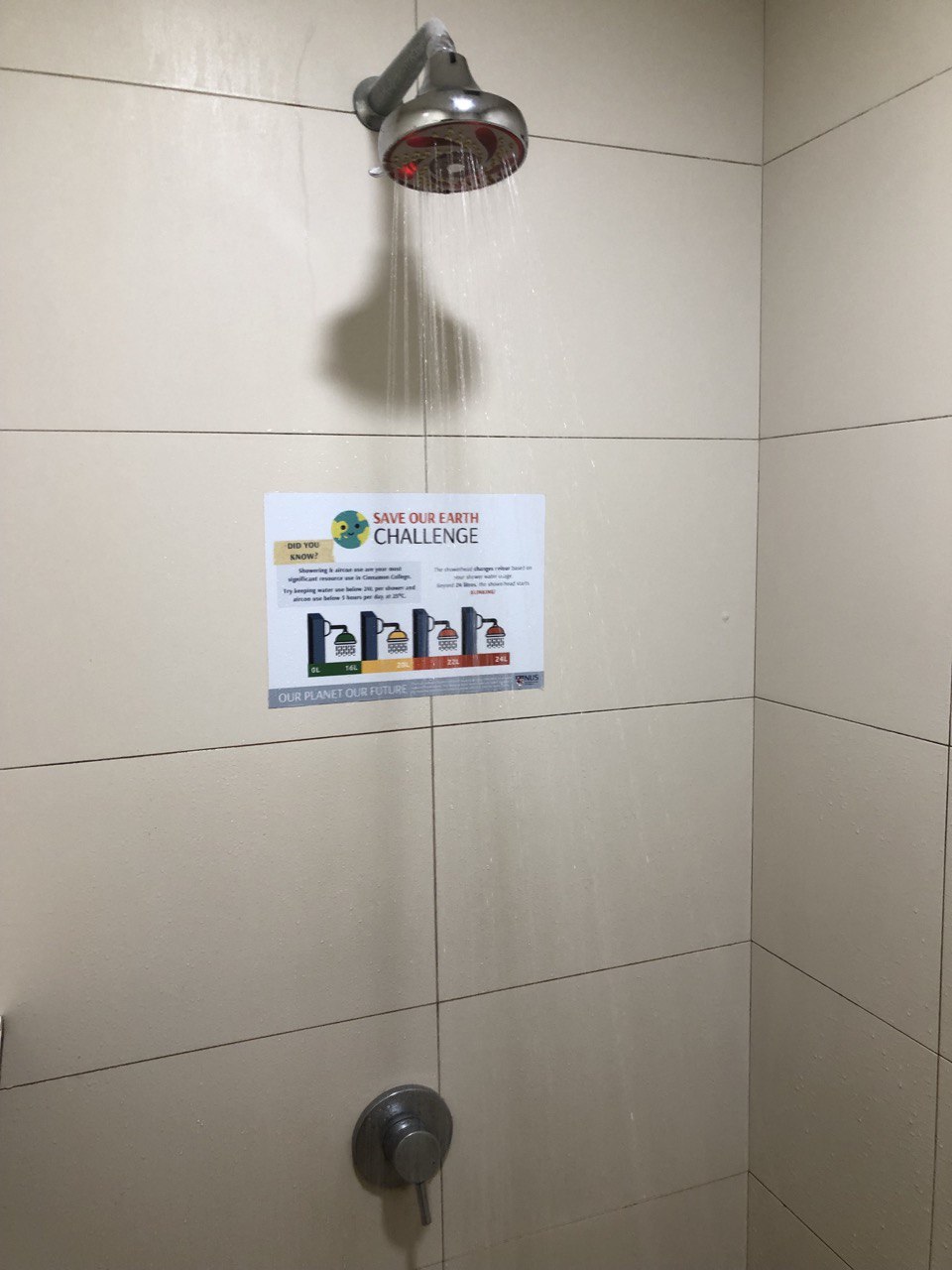}
	\end{subfigure} 
 % \\ \vspace{0.2cm}
	\end{center}
	\begin{footnotesize} 
	\emph{Notes.} At the start of each shower event, the smart shower head displays a green light, which remains on for water use up to 14 litres. As water use increases, the light changes sequentially to yellow, orange, and red at the predetermined thresholds shown above. Beyond 24 litres, the shower head displays a blinking red light. 
	\end{footnotesize}
\end{figure}

\newpage
\subsection{Temperature and Resource Use}

\begin{figure}[H]
	\centering
\caption{Shower Water Usage and Mean Temperature}
 \vspace{1.5em}
	\begin{subfigure}{0.49\textwidth}
    \caption{Intensive margin} 	
  \includegraphics[width=1.01\textwidth]{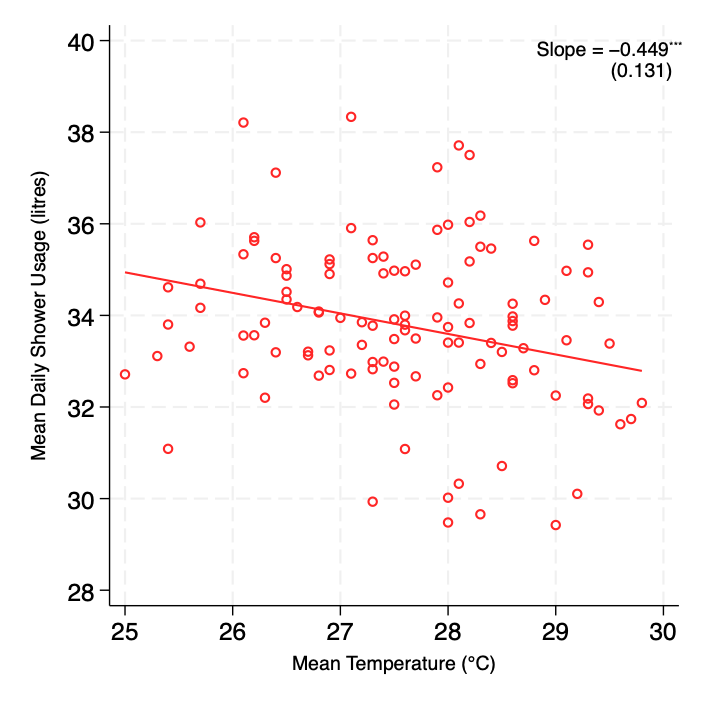} 
  \label{fig:intensive_shower} 
  % \vspace{0.5em}
	\end{subfigure}
	\begin{subfigure}{0.49\textwidth}
    \caption{Extensive margin}
  \includegraphics[width=1.01\textwidth]{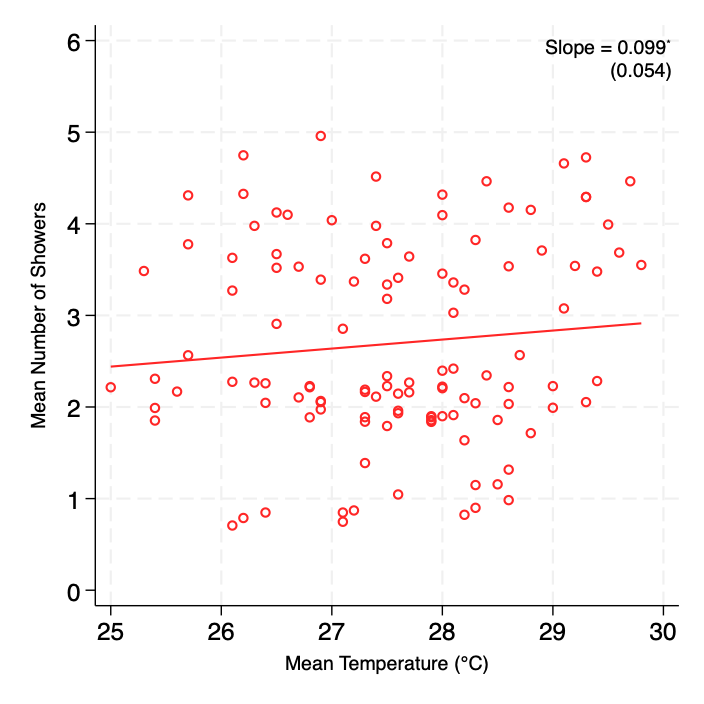} 
  \label{fig:extensive_shower} 
  
	\end{subfigure}
	\begin{minipage}{15cm}
		% \vspace{0.5em}
		\scriptsize \singlespacing{\emph{Notes.} This figure shows the relationship between daily showering use and mean temperature in Singapore. Panel (a) shows the intensive margin, while panel (b) shows the extensive margin. Each observation is a day during the sample period between 10 August 2021 and 4 December 2021. Solid lines show fitted linear relationships.} 
	\end{minipage}
\label{fig:scatterplot_shower} 
\end{figure}

\newpage
\begin{figure}[H]
	\centering
\caption{Air-Conditioning Usage and Mean Temperature}
 \vspace{1.5em}
	\begin{subfigure}{0.49\textwidth}
    \caption{Intensive margin} 	
  \includegraphics[width=1.01\textwidth]{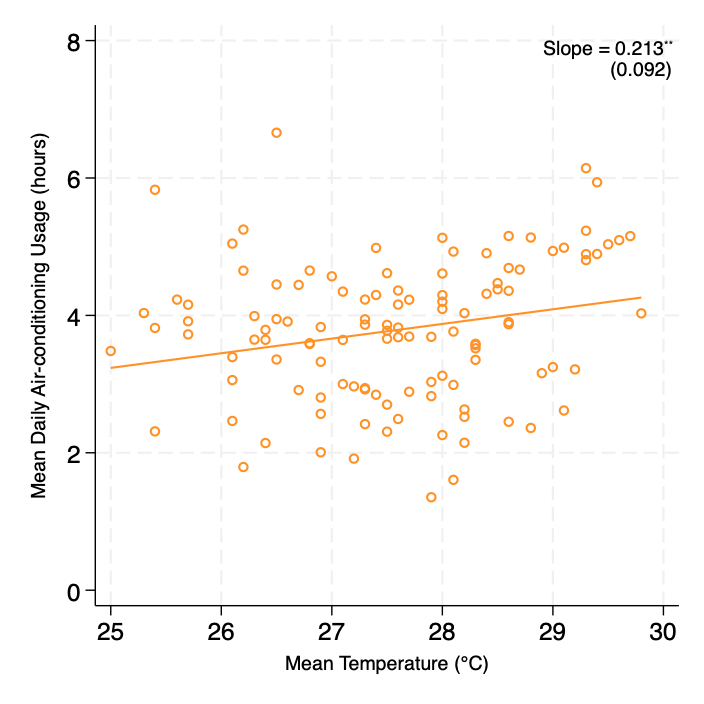} 
  \label{fig:intensive_aircon} 
  % \vspace{0.5em}
	\end{subfigure}
	\begin{subfigure}{0.49\textwidth}
    \caption{Extensive margin} 
  \includegraphics[width=1.01\textwidth]{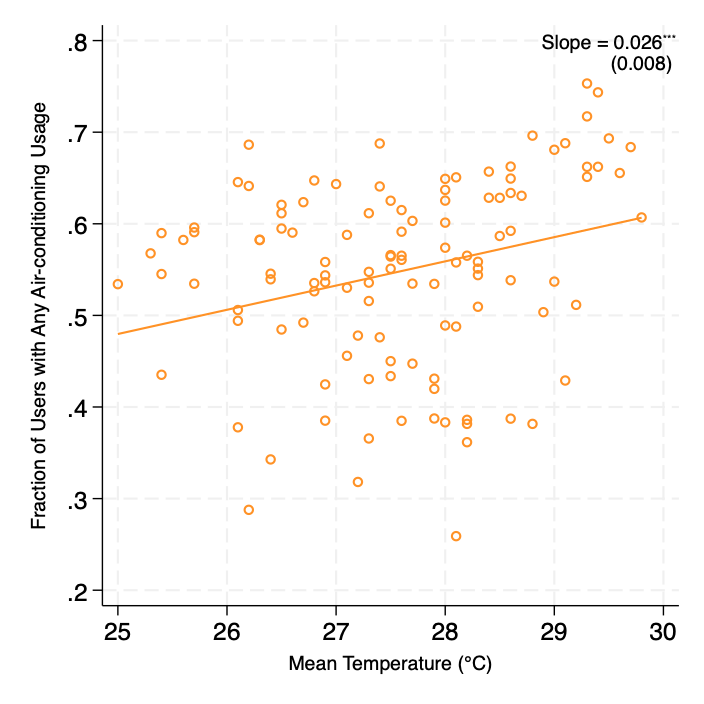} 
  \label{fig:extensive_aircon} 
  
  % \vspace{0.5em}
	\end{subfigure}
	\begin{minipage}{15cm}
		% \vspace{0.5em}
		\scriptsize \singlespacing{\emph{Notes.} This figure shows the relationship between daily air-conditioning use and mean temperature in Singapore. Panel (a) shows the intensive margin, while panel (b) shows the extensive margin. Each observation is a day during the sample period between 10 August 2021 and 4 December 2021. Solid lines show fitted linear relationships.} 
	\end{minipage}
\label{fig:scatterplot_aircon} 
\end{figure}

\newpage
\subsection{Treatment Dynamics}
\begin{figure}[H]
	\centering
 % \vspace{1.5em}
\caption{Temporal Dynamics of Treatment Effects on Targeted Behavior}
\label{fig:temporal_direct} 
 \vspace{1.5em}

    \begin{subfigure}{0.48\textwidth}
        \caption{RTF intervention}
        \includegraphics[width=1.01\textwidth]{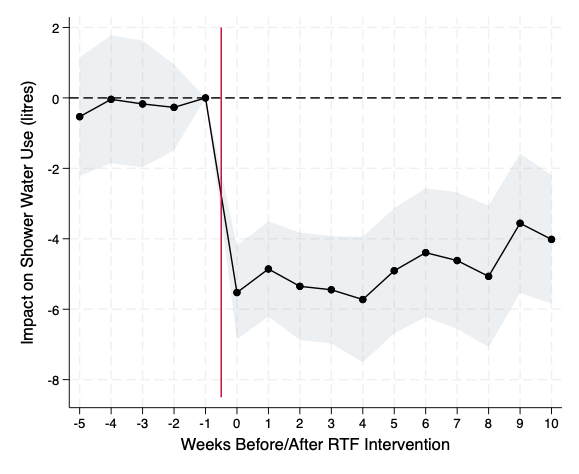} 
        % \label{fig:type1} 
        \vspace{0.5em}
    \end{subfigure}
    
    \vspace{0.5em}
    \begin{subfigure}{0.48\textwidth}
        \caption{SC-S intervention}
        \includegraphics[width=1.01\textwidth]{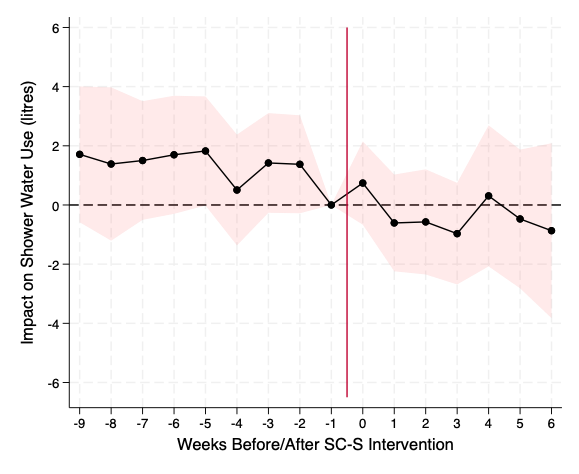} 
        % \label{fig:type3} 
        \vspace{0.5em}
    \end{subfigure}
    \begin{subfigure}{0.48\textwidth}
        \caption{SC-A intervention}
        \includegraphics[width=1.01\textwidth]{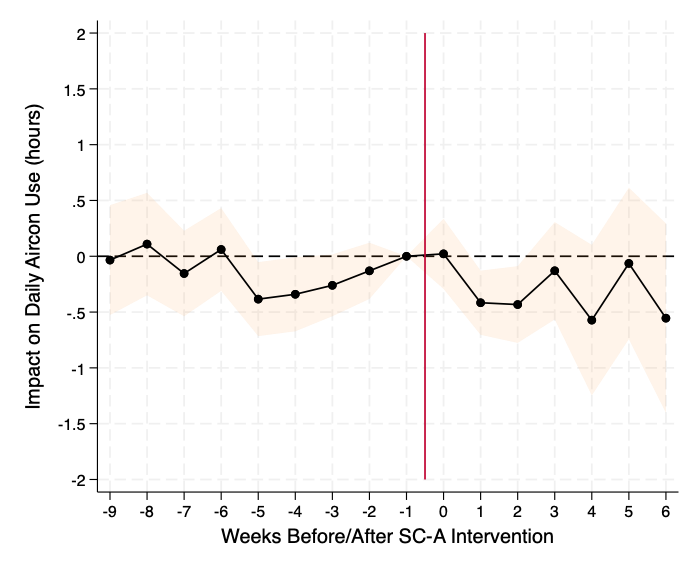} 
        % \label{fig:type4} 
        \vspace{0.5em}
    \end{subfigure}

\begin{minipage}{15cm}
    % \vspace{0.5em}
    \scriptsize \singlespacing{\emph{Notes.} This figure plots the event-time estimates of the direct effects of the RTF and SC-S interventions on shower water use and the direct effects of the SC-A intervention on daily air-conditioning use, by week relative to the week before implementation of the respective interventions (i.e., event time = -1). The estimates are from OLS regressions that interact treatment assignment to RTF, SC-S and SC-A with weekly event-time indicators and include device/room, date and event week fixed effects. The 95\% confidence bands around the estimates are based on standard errors clustered at the bathroom level.} 
\end{minipage}

\vspace{1.0em}
\end{figure}

\begin{figure}[H]
	\centering
 % \vspace{1.5em}
\caption{Temporal Dynamics of Treatment Effects on Non-Targeted Behavior}
\label{fig:temporal_spillover} 
 \vspace{1.5em}

    \begin{subfigure}{0.48\textwidth}
        \caption{RTF intervention}
        \includegraphics[width=1.01\textwidth]{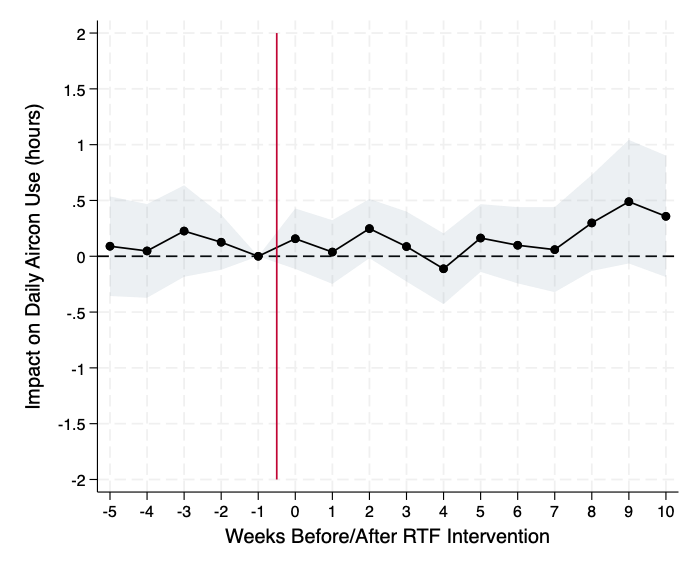} 
        \vspace{0.5em}
    \end{subfigure}
    
    \vspace{0.5em}
    \begin{subfigure}{0.48\textwidth}
        \caption{SC-S intervention}
        \includegraphics[width=1.01\textwidth]{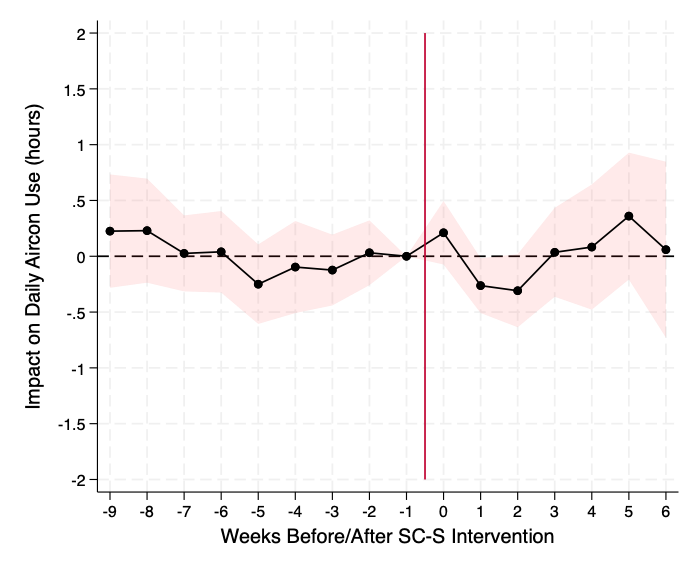} 
        \vspace{0.5em}
    \end{subfigure}
    \begin{subfigure}{0.48\textwidth}
        \caption{SC-A intervention}
        \includegraphics[width=1.01\textwidth]{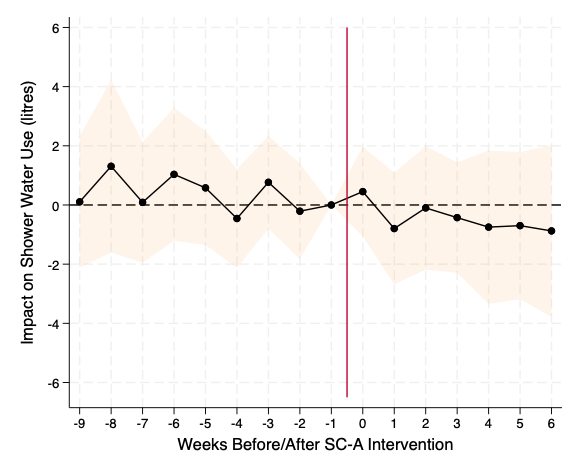} 
        \vspace{0.5em}
    \end{subfigure}

\begin{minipage}{15cm}
    % \vspace{0.5em}
    \scriptsize \singlespacing{\emph{Notes.} This figure plots the event-time estimates of the spillover effects of the RTF and SC-S interventions on daily air-conditioning use and the spillover effects of the SC-A intervention on shower water use, by week relative to the week before implementation of the respective interventions (i.e., event time = -1). The estimates are from OLS regressions that interact treatment assignment to RTF, SC-S and SC-A with weekly event-time indicators and include device/room, date and event week fixed effects. The 95\% confidence bands around the estimates are based on standard errors clustered at the bathroom level.} 
\end{minipage}

\vspace{1.0em}
\end{figure}
\clearpage

\subsection{Assessing Sorting Across Bathrooms}

\begin{figure}[H]
	\centering
\caption{Treatment Effects by Floor Type}
\label{fig:effects_floorType} 
 \vspace{1.5em}
 
	\begin{subfigure}{0.49\textwidth}
    \caption{Shower water usage} 
  \includegraphics[width=1.01\textwidth]{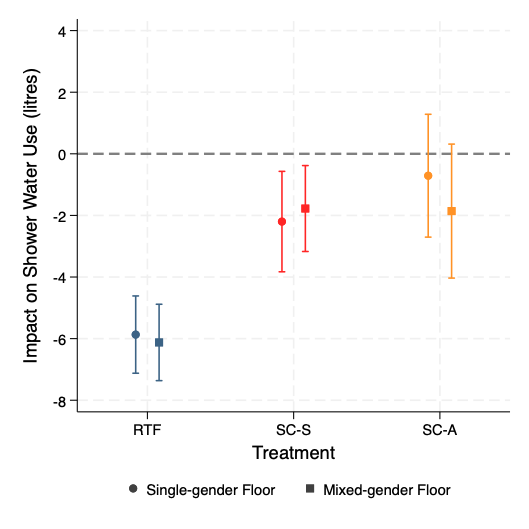} 
	\end{subfigure}
	\begin{subfigure}{0.49\textwidth}
		\caption{Air-conditioning usage} 
  \includegraphics[width=1.01\textwidth]{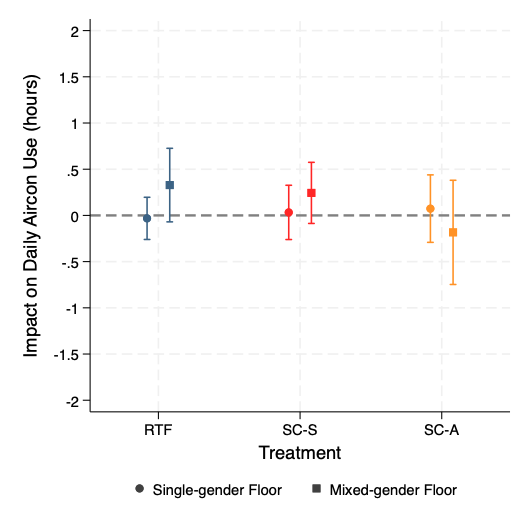} 
	\end{subfigure}

    \begin{minipage}{15cm}
    % \vspace{0.5em}
    \scriptsize \singlespacing{\emph{Notes.} This figure plots the coefficient estimates of the direct and spillover effects of the RTF, SC-S, and SC-A interventions by floor type. Panel (a) uses shower water usage as the dependent variable, while panel (b) uses air-conditioning usage as the dependent variable. The estimates are from OLS regressions that interact treatment indicators for RTF, SC-S, and SC-A with indicators for single-gender and mixed-gender floors, and include device/room and date fixed effects. The 95\% confidence bands around the estimates are based on standard errors clustered at the bathroom level.}     
\end{minipage}

\end{figure}

To assess the concern around sorting across bathrooms, we test whether treatment effects differ between single-gender and mixed-gender floors. Switching to an alternative bathroom is more costly on mixed-gender floors, as residents would need to travel farther (e.g., to a different floor). For instance, if residents who dislike the interventions and would otherwise respond most strongly sort into control bathrooms, the estimated treatment effects would be biased toward zero. More generally, if sorting were driving our results, we would expect treatment effects to differ across floor types. Reassuringly, we find no statistically significant differences in treatment effects across floor types, which alleviates concerns about sorting. Figure \ref{fig:effects_floorType} plots the estimated treatment effects by floor type.

\bigskip

\clearpage
\subsection{Additional Heterogeneity Analyses}

This subsection reports additional heterogeneity analyses by baseline usage, daily temperature, and room type.

% These analyses are used to assess whether the estimated null effects on air-conditioning usage mask meaningful heterogeneity along observable margins.

\bigskip

\begin{figure}[H]
	\centering
\caption{Treatment Effects by Baseline Use Quartile}
\label{fig:treatment_baseline_quartile}
 \vspace{1.5em}
 
	\begin{subfigure}{0.49\textwidth}
    \caption{Shower water usage}
  \includegraphics[width=1.01\textwidth]{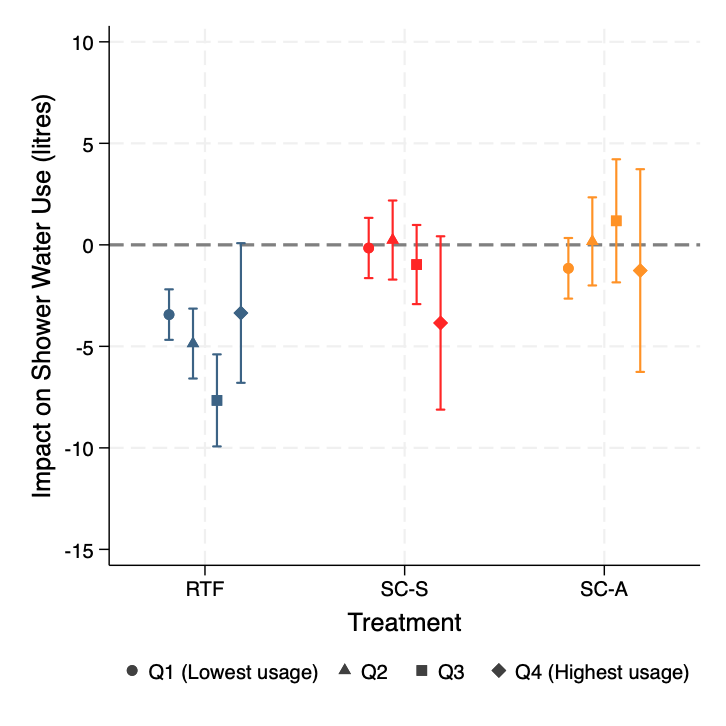} 
	\end{subfigure}
	\begin{subfigure}{0.49\textwidth}
    \caption{Air-conditioning usage} 
  \includegraphics[width=1.01\textwidth]{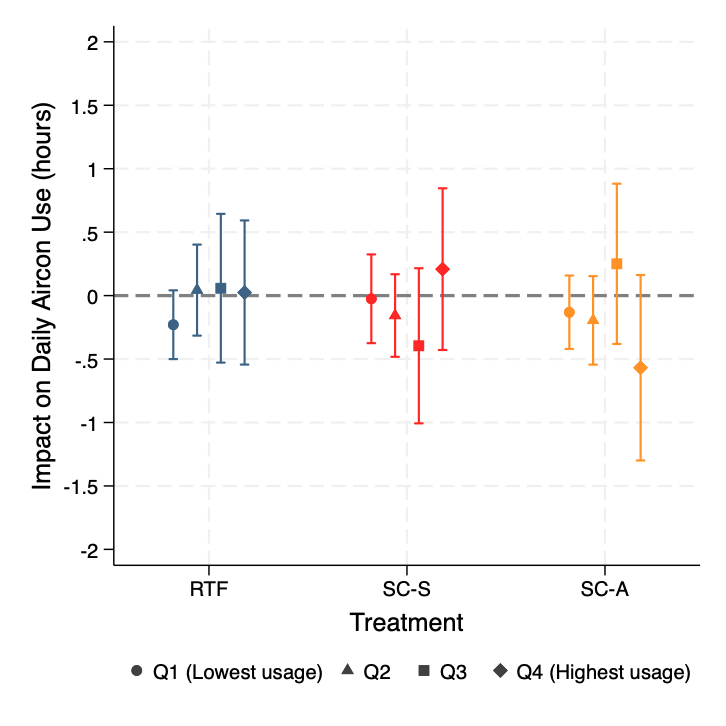} 
	\end{subfigure}
    
    \begin{minipage}{15cm}
    % \vspace{0.5em}
    \scriptsize \singlespacing{\emph{Notes.} This figure plots the coefficient estimates of the direct and spillover effects of the RTF, SC-S, and SC-A interventions by baseline use quartile. Panel (a) uses shower water usage as the dependent variable, while panel (b) uses air-conditioning usage as the dependent variable. The estimates are from OLS regressions that interact treatment indicators for RTF, SC-S, and SC-A with indicators for each baseline use quartile (Q1--Q4), and include device/room and date fixed effects. The 95\% confidence bands around the estimates are based on standard errors clustered at the bathroom level.}     
\end{minipage}

\end{figure}
\clearpage

\newpage
\begin{figure}[H]
	\centering
 % \vspace{1.5em}
\caption{Treatment Effects on Targeted Behavior by Daily Temperature}
\label{fig:temperature_direct} 
 \vspace{1.5em}

    \begin{subfigure}{0.48\textwidth}
        \caption{RTF intervention}
        \includegraphics[width=1.01\textwidth]{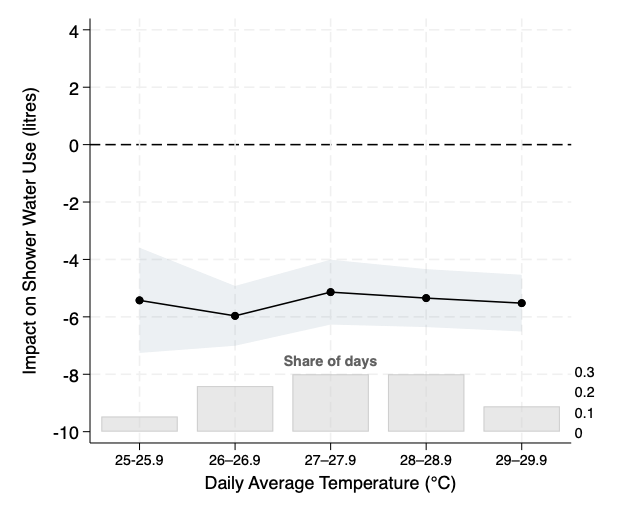} 
        \vspace{0.5em}
    \end{subfigure}
    
    \vspace{0.5em}
    \begin{subfigure}{0.48\textwidth}
        \caption{SC-S intervention}
        \includegraphics[width=1.01\textwidth]{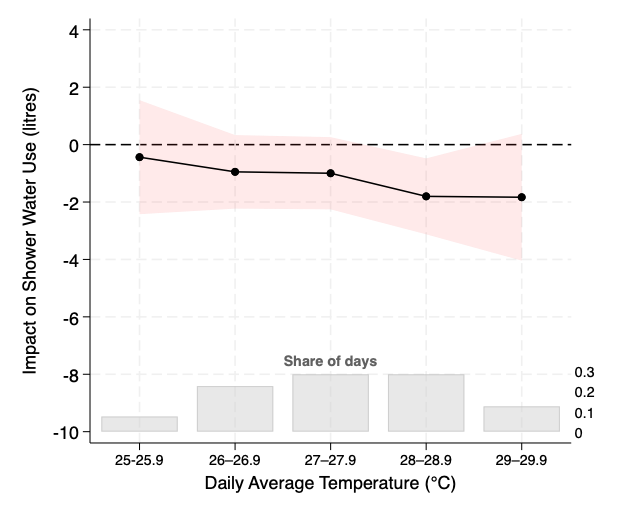} 
        \vspace{0.5em}
    \end{subfigure}
    \begin{subfigure}{0.48\textwidth}
        \caption{SC-A intervention}
        \includegraphics[width=1.01\textwidth]{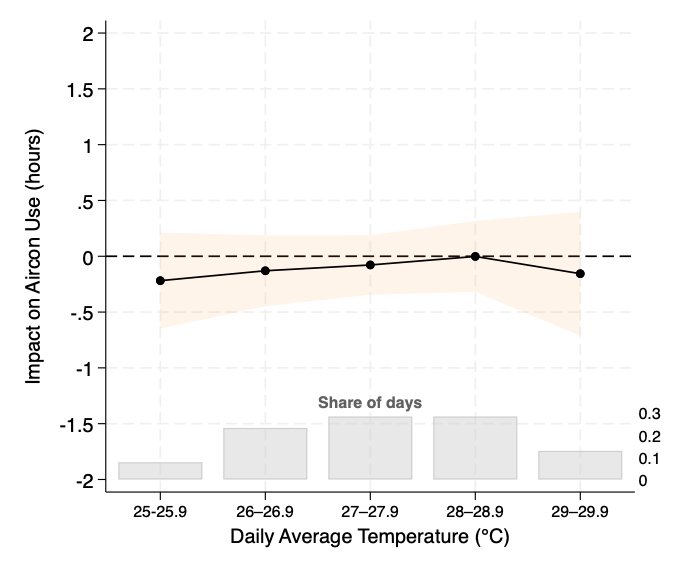} 
        \vspace{0.5em}
    \end{subfigure}

\begin{minipage}{15cm}
    % \vspace{0.5em}
    \scriptsize \singlespacing{\emph{Notes.} This figure plots the coefficient estimates of the direct effects of the RTF and SC-S interventions on shower water use and the direct effects of the SC-A intervention on daily air-conditioning use, by temperature bin. The estimates are from OLS regressions that interact treatment assignment to RTF, SC-S and SC-A with indicators for temperature bins (defined in 1°C intervals: 25.0–25.9°C, 26.0–26.9°C, ..., 29.0–29.9°C), and include device/room and date fixed effects. The 95\% confidence bands around the estimates are based on standard errors clustered at the bathroom level. Histograms at the bottom display the share of days in each temperature bin during the experimental period.} 
\end{minipage}

\end{figure}

\begin{figure}[H]
	\centering
 % \vspace{1.5em}
\caption{Treatment Effects on Non-Targeted Behavior by Daily Temperature}
\label{fig:temperature_spillover} 
 \vspace{1.5em}

    \begin{subfigure}{0.48\textwidth}
        \caption{RTF intervention}
        \includegraphics[width=1.01\textwidth]{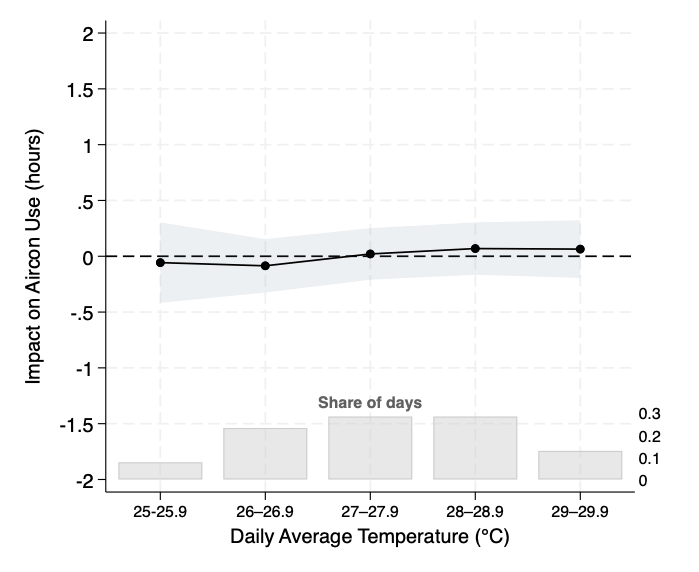} 
        \vspace{0.5em}
    \end{subfigure}
    
    \vspace{0.5em}
    \begin{subfigure}{0.48\textwidth}
        \caption{SC-S intervention}
        \includegraphics[width=1.01\textwidth]{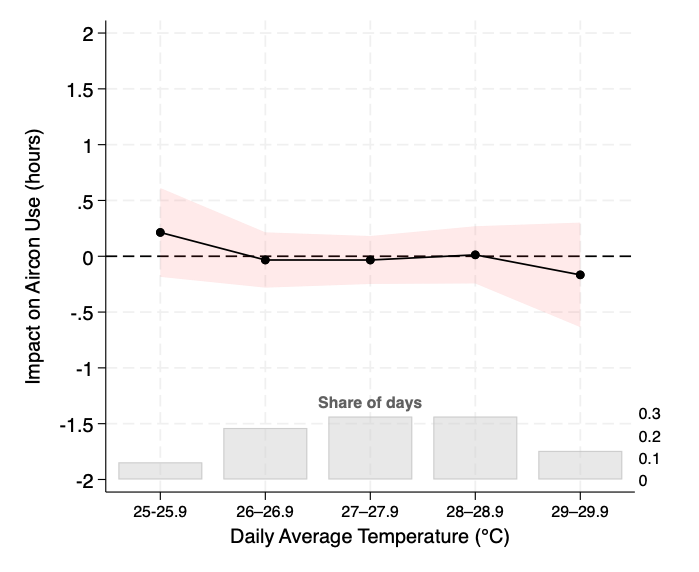} 
        \vspace{0.5em}
    \end{subfigure}
    \begin{subfigure}{0.48\textwidth}
        \caption{SC-A intervention}
        \includegraphics[width=1.01\textwidth]{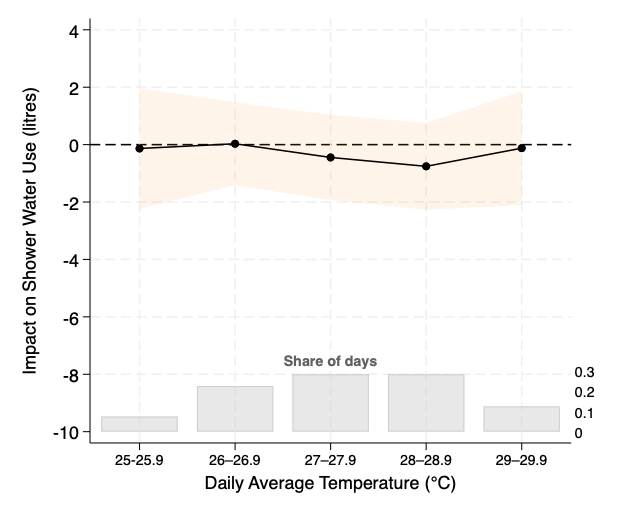} 
        \vspace{0.5em}
    \end{subfigure}

\begin{minipage}{15cm}
    % \vspace{0.5em}
    \scriptsize \singlespacing{\emph{Notes.} This figure plots the coefficient estimates of the spillover effects of the RTF and SC-S interventions on daily air-conditioning use and the spillover effects of the SC-A intervention on shower water use, by temperature bin. The estimates are from OLS regressions that interact treatment assignment to RTF, SC-S and SC-A with indicators for temperature bins (defined in 1°C intervals: 25.0–25.9°C, 26.0–26.9°C, ..., 29.0–29.9°C), and include device/room and date fixed effects. The 95\% confidence bands around the estimates are based on standard errors clustered at the bathroom level. Histograms at the bottom display the share of days in each temperature bin during the experimental period.} 
\end{minipage}

\end{figure}

\begin{figure}[H]
	\centering
\caption{Treatment Effects by Room Type}
\label{fig:treatment_room_type}
 \vspace{1.5em}
 
	\begin{subfigure}{0.49\textwidth}
    \caption{Shower water usage}
  \includegraphics[width=1.01\textwidth]{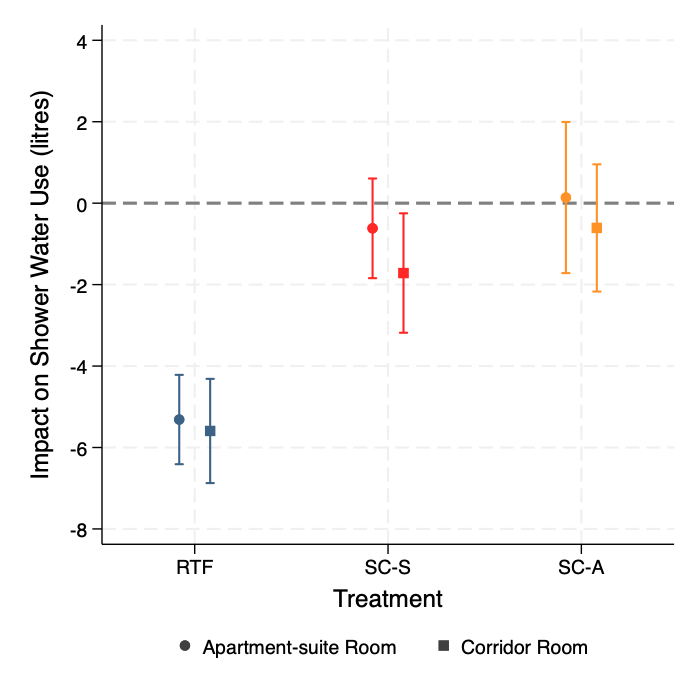} 
	\end{subfigure}
	\begin{subfigure}{0.49\textwidth}
    \caption{Air-conditioning usage}
  \includegraphics[width=1.01\textwidth]{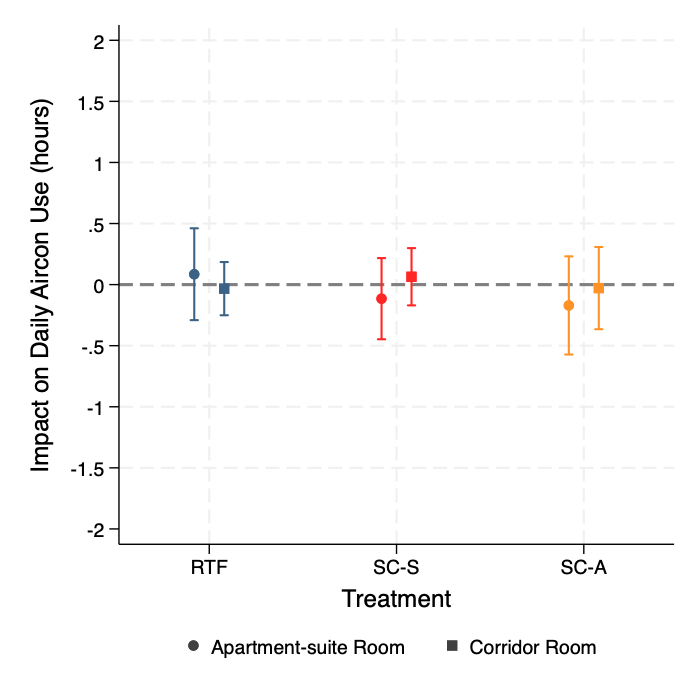} 
	\end{subfigure}

        \begin{minipage}{15cm}
    % \vspace{0.5em}
    \scriptsize \singlespacing{\emph{Notes.} This figure plots the coefficient estimates of the direct and spillover effects of the RTF, SC-S, and SC-A interventions by room type. Panel (a) uses shower water usage as the dependent variable, while panel (b) uses air-conditioning usage as the dependent variable. The estimates are from OLS regressions that interact treatment indicators for RTF, SC-S, and SC-A with indicators for 6-bedroom apartment-suite rooms and single corridor rooms, and include device/room and date fixed effects. The 95\% confidence bands around the estimates are based on standard errors clustered at the bathroom level.}     
\end{minipage}

\end{figure}

\newpage
\subsection{Randomization Checks}

Table \ref{tab:sum_stats} presents the mean baseline characteristics for bathrooms and bedrooms across our six experimental conditions. In particular, the two main observables (i.e., baseline shower water usage and air-conditioning usage) are comparable across all conditions, with no statistically significant differences.

{\singlespacing

\begin{table}[H]
    \centering 
	% \caption{Mean baseline characteristics by experimental condition} 
    	\caption{Sample and Balance} 
    \label{tab:sum_stats}
	\begin{threeparttable}
 \resizebox{\textwidth}{!}{% Resize table to fit within the textwidth of the page
    \setlength{\tabcolsep}{12pt} % Increase column spacing
		\begin{tabular}{l*{7}{c}}
			\toprule
			\toprule
			 & \multicolumn{7}{c}{Experimental Condition} \\ 
            \cmidrule(lr){2-8} 
            % & T1 & T2 & T3 & T4 & T5 & T6 \\ 
            & \footnotesize{Control} & \footnotesize{SC-S} & \footnotesize{SC-A} & \footnotesize{RTF} & \footnotesize{RTF+SC-S} & \footnotesize{RTF+SC-A} & \footnotesize{p-value} \\ 
            & (1) & (2) & (3) & (4) & (5) & (6) & (7) \\ 
			\midrule \addlinespace 
			Shower water usage (litres)    &   37.23 &  36.44 & 38.08  & 39.74 & 37.03 &  35.47 & 0.764  \\ \addlinespace
			Air-conditioning usage (hours) &   3.61 & 3.66  &   3.72    & 3.66 &  3.41 & 3.54 & 0.959 \\ \addlinespace
			Share of suites  &  0.56 & 0.59 & 0.51 & 0.52 & 0.58 & 0.51  & 0.971 \\ \addlinespace
            Floor & 10.59  & 12.87 & 11.32  & 10.96 &  10.50 & 10.21  & 0.170 \\ 
			\midrule \addlinespace
			No. of shower heads & 106 & 73 & 69 & 91 & 71 & 71 & -- \\  
            No. of bathrooms (clusters) & 54 & 39 & 37 & 48 & 40 & 39 & -- \\  
            No. of air-conditioned bedrooms  & 192 & 231 & 219 & 212 & 220 & 202 & -- \\
			\bottomrule
		\end{tabular}
  }
		\begin{tablenotes}
			\scriptsize\vspace{0.1cm} 
			\item \emph{Notes.} 
            Each p-value is from an F-test of joint significance in an OLS regression of the variable on treatment group dummies.
            % This table presents the mean baseline characteristics for bathrooms and bedrooms across our six experimental conditions. All key observables (baseline shower water usage and air conditioning usage in particular) are comparable across all conditions, with no statistically significant differences detected.
		\end{tablenotes}
	\end{threeparttable}	
\end{table}
}

\newpage
\subsection{Treatment Effects on the Extensive Margin}

Table \ref{tab:appendix} illustrates that the number of showers taken and the fraction of residents using air conditioning each day do not vary across experimental conditions. This provides evidence against residents responding to any of the interventions on the extensive margin.

%%%%%%%%%%%%%
%% Table 4 %%
%%%%%%%%%%%%%
{\singlespacing

% \begin{sidewaystable}
\begin{table}[H]
\centering 
	\caption{Effects of Real-Time Feedback and Social Comparisons (Extensive Margin)} 
    \label{tab:appendix}
	\begin{threeparttable}
 \resizebox{\textwidth}{!}{% Resize table to fit within the textwidth of the page
    % \setlength{\tabcolsep}{12pt} % Increase column spacing
		% \begin{tabular}{l*{4}{c}}
        \begin{tabular}{l*{4}{>{\centering\arraybackslash}p{2.2cm}}}
			\toprule
			\toprule
			Dependent variable: & \multicolumn{2}{c}{Number of showers per day} & \multicolumn{2}{c}{Fraction using aircon each day} \\
            \cmidrule(lr){2-3} \cmidrule(lr){4-5} 
            & (1) & (2) & (3) & (4) \\ 
			\midrule \addlinespace 
			Real-Time Feedback (RTF)    &       0.225   &       0.304 &       0.003   &       0.005   \\
			 &     (0.226)   &     (0.222)  &     (0.013)   &     (0.014)   \\ \addlinespace
			Social Comparison for Shower (SC-S) &      --0.287   &      --0.160  &       0.005   &       0.010   \\
            &     (0.204)   &     (0.264)  &     (0.013)   &     (0.016)   \\ \addlinespace
            Social Comparison for Aircon (SC-A) &      --0.292   &      --0.108  &      --0.002   &      --0.001   \\
			&     (0.237)   &     (0.318) &     (0.014)   &     (0.018)   \\ \addlinespace
            SC-S $\times$ RTF &               &      --0.267 &               &      --0.010   \\
            &               &     (0.314) &               &     (0.020)   \\ \addlinespace
            SC-A $\times$ RTF &               &      --0.370 &               &      --0.003   \\
            &               &     (0.393) &               &     (0.023)   \\ \addlinespace
            \addlinespace
            Baseline Mean &       4.789 &       4.789 &       0.532&       0.532 \\
             &     (4.339)   &     (4.339)  &     (0.499)   &     (0.499)   \\ 
             \addlinespace  
			\midrule \addlinespace
			Device/Room FEs & \ding{51} & \ding{51} & \ding{51} & \ding{51}   \\ \addlinespace
			Date FEs & \ding{51} & \ding{51} & \ding{51} & \ding{51}  \\ \addlinespace
			\(R^{2}\) & 0.343 & 0.343 & 0.435 & 0.435 \\ \addlinespace  
			Observations & 57,953 & 57,953 & 147,376 & 147,376 \\   
			\bottomrule
		\end{tabular}
  }
		\begin{tablenotes}
			\scriptsize\vspace{0.1cm} 
            \parbox{.95\textwidth}{
			\item \emph{Notes.} This table reports OLS estimates of the treatment effects of the RTF, SC-S, and SC-A interventions on the extensive margin. All specifications include device/room and day fixed effects. Columns (1) and (2) report estimates for the number of showers per day, while columns (3) and (4) report estimates for the fraction of rooms using air conditioning each day. Standard errors clustered at the bathroom level in parentheses. 
			\item * $p<0.10$, ** $p<0.05$, *** $p<0.01$
            }
		\end{tablenotes}
	\end{threeparttable}	
\end{table}
% \end{sidewaystable}
}

\newpage
\setcounter{table}{0} 
\setcounter{figure}{0}
\renewcommand{\thetable}{B\arabic{table}}
\renewcommand{\thefigure}{B\arabic{figure}}

\section{Details on Performance Ranking Measures}

\subsection{Construction of the Percentile Rankings}

As detailed in the main text, each residential college features two room types: single corridor rooms and single rooms within shared apartment suites. Residents in corridor rooms use common bathrooms, while residents in apartment suites use the private bathroom associated with their suite. We therefore construct performance rankings at the bathroom level. For shower usage, each bathroom is ranked based on the average shower water usage of residents using that bathroom. For air-conditioning usage, each bathroom is ranked based on the average air-conditioning usage of the rooms linked to that bathroom. In both cases, the comparisons are made among bathrooms in the same residential college and room type. 

The timing of the ranking construction follows the social comparison posters. Specifically, the posters were introduced at the beginning of phase 2 and refreshed midway through the phase. The initial posters used average usage over the two-week window immediately preceding phase 2. When the posters were refreshed midway through phase 2, the rankings were recomputed using average usage over the preceding two-week window. This yields two ranking measures: an initial ranking and a refreshed ranking.

The rankings shown on the posters are ordinal ranks, with rank 1 corresponding to the top-performing bathroom (i.e., lowest usage) in the comparison group. We convert these rankings into percentiles from 0 to 100, then normalize them to range from $-50$ to 50, where 0 corresponds to the median-ranked bathroom. We orient the percentile measure so that higher values correspond to more favorable relative performance, i.e., lower usage within the relevant comparison group. 

% For example, Tembusu College has 34 corridor bathrooms and 41 suite bathrooms. The ranking of a bathroom (e.g., 5th out of 34 corridor bathrooms) is then converted into a percentile, ranging from 0 to 100. To ease interpretation in our regression analyses, we normalize these percentiles to a scale from $-$50 to 50, where a value of 0 indicates the median usage. These measures are denoted by $\text{Percentile}_{Shower}$ and $\text{Percentile}_{Aircon}$ for shower water usage and air-conditioning usage, respectively, in Table \ref{tab:main2}.

\subsection{Robustness: Fixed and Time-Varying Percentile Rankings}

For the heterogeneity analysis in Table \ref{tab:main2}, our specification uses only the initial percentile ranking, constructed before the introduction of social comparisons, and holds it fixed throughout the analysis. These bathroom-level measures are denoted by $\text{PercentileRank}_{i}^{S}$ and $\text{PercentileRank}_{i}^{A}$ for shower water usage and air-conditioning usage, respectively.

This is our preferred specification because the percentile rankings are predetermined relative to the onset of the social comparison interventions, so the interaction effects are not mechanically affected by treatment responses during phase 2. As a robustness check, we also estimate an alternative specification using time-varying percentile rankings, $\text{PercentileRank}_{it}^{S}$ and $\text{PercentileRank}_{it}^{A}$, which incorporate the refreshed rankings midway through phase 2. The estimates are very similar to those reported in Table \ref{tab:main2} (results available upon request). 

\begin{figure}[H]
	\centering
	\caption{Correlation Between Initial and Refreshed Percentile Rankings}
	\vspace{1.5em}
	\begin{subfigure}{0.49\textwidth}
		\caption{Shower usage}
		\includegraphics[width=1.01\textwidth]{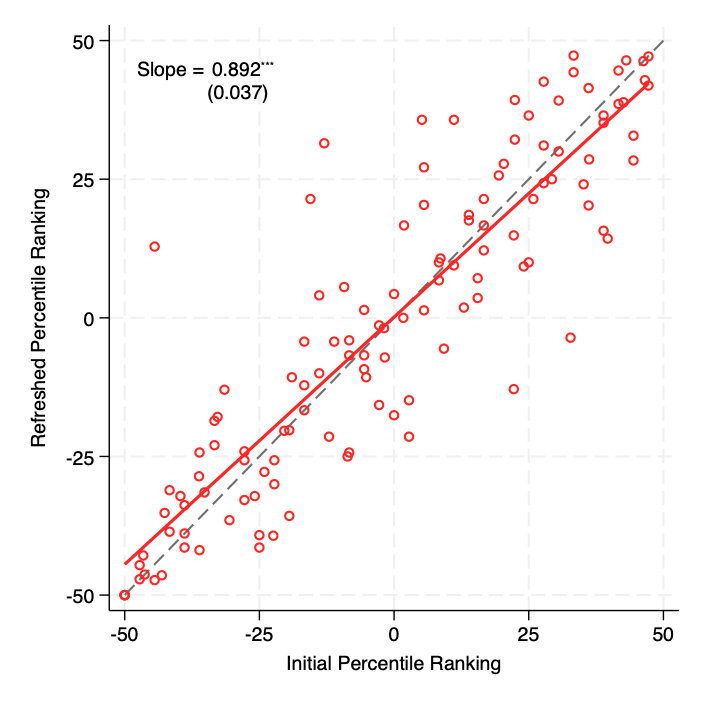}
		\label{fig:percentile_corr_shower}
	\end{subfigure}
	\begin{subfigure}{0.49\textwidth}
		\caption{Air-conditioning usage}
		\includegraphics[width=1.01\textwidth]{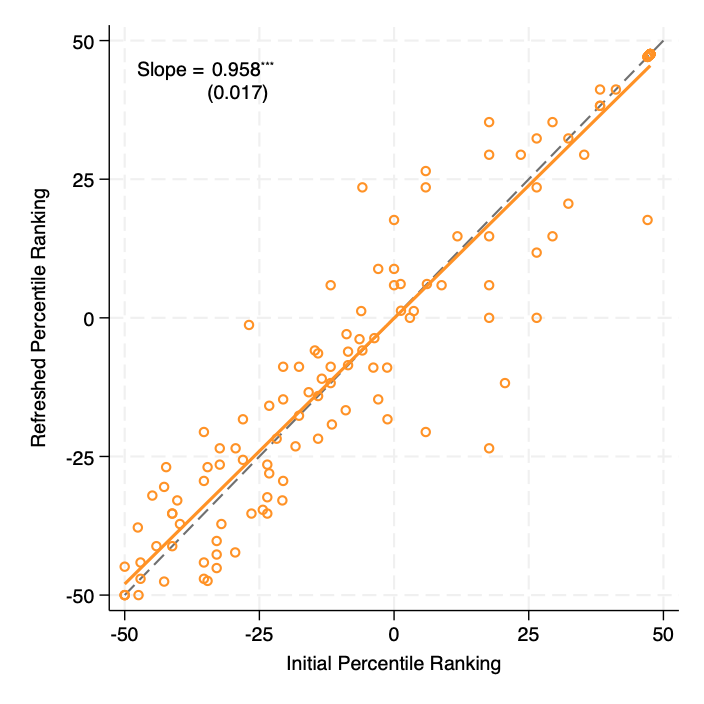}
		\label{fig:percentile_corr_aircon}
	\end{subfigure}
	\begin{minipage}{15cm}
		\scriptsize \singlespacing{\emph{Notes.} This figure plots each bathroom's initial percentile ranking against its refreshed percentile ranking. Panel (a) shows shower usage, and panel (b) shows air-conditioning usage. Initial rankings are constructed based on usage in the two-week window immediately preceding phase 2. Refreshed rankings are constructed when the social comparison posters were refreshed midway through phase 2, based on usage in the preceding two-week window. Percentile rankings are computed within each residential college and bathroom type, then normalized to range from $-50$ to 50. Higher values indicate more favorable rankings, corresponding to lower usage within the relevant comparison group. Solid lines show fitted linear relationships, and dashed gray lines indicate 45-degree lines.}
	\end{minipage}
	\label{fig:percentile_corr}
\end{figure}

This is consistent with the strong correlation between the initial and refreshed percentile rankings. Figure \ref{fig:percentile_corr} plots the relationship between the initial and refreshed percentile rankings for shower water usage and air-conditioning usage, respectively. The correlations are 0.89 for shower rankings and 0.96 for air-conditioning rankings, indicating that the percentile rankings are highly correlated over time.

\stopcontents[appendix]
                           
\end{document}